\documentclass[12pt]{article}
\usepackage[dvips]{graphicx}
\usepackage{tabularx}

\makeatletter
\newtoks\@stequation

\def\subequations{\refstepcounter{equation}%
  \edef\@savedequation{\the\c@equation}%
  \@stequation=\expandafter{\theequation}
  \edef\@savedtheequation{\the\@stequation}
  \edef\oldtheequation{\theequation}%
  \setcounter{equation}{0}%
  \def\theequation{\oldtheequation\alph{equation}}}

\def\endsubequations{%
  \ifnum\c@equation < 2 \@warning{Only \the\c@equation\space subequation
    used in equation \@savedequation}\fi
  \setcounter{equation}{\@savedequation}%
  \@stequation=\expandafter{\@savedtheequation}%
  \edef\theequation{\the\@stequation}%
  \global\@ignoretrue}

\def\eqnarray{\stepcounter{equation}\let\@currentlabel\theequation
\global\@eqnswtrue\m@th
\global\@eqcnt\z@\tabskip\@centering\let\\\@eqncr
$$\halign to\displaywidth\bgroup\@eqnsel\hskip\@centering
     $\displaystyle\tabskip\z@{##}$&\global\@eqcnt\@ne
      \hfil$\;{##}\;$\hfil
     &\global\@eqcnt\tw@ $\displaystyle\tabskip\z@{##}$\hfil
   \tabskip\@centering&\llap{##}\tabskip\z@\cr}
\makeatother

\def\cerenkov{$\check{\rm C}$erenkov~}
\def\mohorovicic{Mohorovi$\check{\rm c}$i$\acute{\rm c}$~}
\def\mutoe{$\nu_{\mu} \to \nu_e$~}

\def\density{{\rm g/cm}^3}
\def\km{\rm km}
\def\deg#1{$#1^{\circ}~$}

\def\dmns{\delta_{\rm \footnotesize MNS}}

\def\lsir{$\sin ^2 2 \theta_{\rm \footnotesize RCT}$~}

\def\PR#1#2#3{Phys. Rev. {\bf #1}, #2 (#3)}

\def\PL#1#2#3{Phys. Lett. {\bf #1}, #2 (#3)}

\def\PTP#1#2#3{Prog. Theor. Phys. {\bf #1}, #2 (#3)}

\def\PRD#1#2#3{Phys. Rev. {\bf D#1},~#2 (#3)}

\def\EPJC#1#2#3{Eur. Phys. J. {\bf C#1} #2 (#3)}
\def\PLB#1#2#3{Phys. Lett. {\bf B#1} #2 (#3)}

\def\eqref#1{eq.~(\ref{#1})}
\def\figref#1{Fig.~\ref{#1}}
\def\tableref#1{Table~\ref{#1}}

\def\etal{{\it et al.}}
\def\ibid{{\it ibid.~}}

\def\simgt{\lower.5ex\hbox{$\; \buildrel > \over \sim \;$}}
\def\simlt{\lower.5ex\hbox{$\; \buildrel < \over \sim \;$}}

\title{The earth matter effects in neutrino oscillation experiments from Tokai to Kamioka and Korea.
}

\author{Kaoru Hagiwara$^{1,~2}$, 
Naotoshi Okamura$^3$\thanks{e-mail:nokamura@yamanashi.ac.jp} , 
and 
Ken-ichi Senda$^{1}$\thanks{e-mail:senda@post.kek.jp}
\\ \\
{\it \small $^1$KEK Theory Center, Tsukuba, 305-0801 Japan} \\ 
{\it \small $^2$Sokendai, (The Graduate University for Advanced Studies),  Tsukuba, 305-0801 Japan}\\
{\it \small $^3$ Faculty of Engineering,  University of Yamanashi, Kofu, Yamanashi, 400-8511, Japan
}
}
\date{}

\begin{document}
\maketitle
\vspace{-9.5cm}
\begin{flushright}
KEK-TH-1100
\hspace*{3ex}
\end{flushright}
\vspace{ 9.5cm}
\vspace{-2.0cm}
\begin{abstract}
We study the earth matter effects in the Tokai-to-Kamioka-and-Korea
experiment (T2KK), which is a proposed extension of the T2K (Tokai-to-Kamioka) neutrino
oscillation experiment between J-PARC at Tokai and Super-Kamiokande (SK) in Kamioka,
where an additional detector is placed in Korea along the same neutrino
beam line.
By using recent geophysical measurements, 
we examine the earth matter effects on the oscillation probabilities at Kamioka
and Korea.
The average matter density along the
Tokai-to-Kamioka baseline is found to be 2.6 $\density$,
and that for the Tokai-to-Korea baseline is 2.85,
2.98, and 3.05 $\density$ for the baseline length of $L$- = 1000, 1100, and 1200 km, respectively.
The uncertainty of the average density is about $6\%$,
which is determined by the
uncertainty in the correlation between the
accurately measured sound velocity and the matter density.
The effect of the matter density distribution along the baseline is studied
by using the step function approximation and the Fourier analysis.
We find that the $\nu_\mu \to \nu_e$ oscillation probability
is dictated mainly by the average matter density,
with small but non-negligible contribution
from the real part of the first Fourier mode.
We also find that the sensitivity of the T2KK
experiment on the neutrino mass hierarchy does not
improve significantly by reducing the matter density error
from $6\%$ to $3\%$,
since the measurement is limited by statistics for
the minimum scenario of T2KK with SK at Kamioka and
a 100 kt detector in Korea considered in this report. 
The sensitivity of the T2KK experiment on the neutrino mass hierarchy
improves significantly by splitting the total beam
time into neutrino and anti-neutrino runs,
because the matter effect term contributes to the oscillation amplitudes
with the opposite sign.
\end{abstract}
\newpage
\section{Introduction}
\hspace{11pt}
Many neutrino oscillation experiments have been performed to measure the
neutrino oscillation parameters.
In the three neutrino model,
there are nine fundamental parameters;
three masses ($m_1$, $m_2$, $m_3$),
three mixing angles ($\theta_{12}$, $\theta_{13}$, $\theta_{23}$)
and
three phases ($\delta$, $\phi_1$, $\phi_2$)
of the lepton-flavor-mixing matrix, MNS matrix \cite{mns}.
Among the three CP phases,
one ($\delta$) is lepton number conserving
and two ($\phi_1$, $\phi_2$) are lepton number non-conserving phases
of the neutrino Majorana masses.
Out of the nine parameters,
the neutrino-flavor oscillation experiments
can measure six parameters;
two mass-squared differences ($m^2_2 - m^2_1$, $m^2_3 - m^2_1$),
all the three mixing angles ($\theta_{12}$, $\theta_{13}$, $\theta_{23}$),
and the lepton number conserving phase ($\delta$).
Among them, 
the magnitude of the larger mass-squared difference,
$|m^2_3 - m^2_1|$, and a mixing angle,
$\theta_{23}$ , 
have been measured by the atmospheric neutrino experiments\cite{sk1998, atomos-result},
which have been confirmed by the
accelerator based long-baseline (LBL)
neutrino oscillation experiments, K2K \cite{k2k}
and MINOS \cite{minos}.
The smaller mass-squared difference, $m^2_2 - m^2_1$,
and the corresponding mixing angle, $\theta_{12}$,
have been measured by the solar neutrino oscillation experiments
\cite{sol-result},
and the KamLAND experiment \cite{kamland}
that observes the oscillation of reactor anti-neutrinos
at more than 100 km distance.
The third mixing angle,
$\theta_{13}$,
has been looked for in the oscillation of
reactor anti-neutrinos at about 1km distance.
No oscillation has been observed and only the upper bound on $\sin^2 2\theta_{13}$
of about 0.14 has been reported \cite{chooz, paloverde}.
\par
Summing up, out of the nine parameters of the
three neutrino model,
six parameters can be observed in neutrino-flavor oscillation
experiments,
among which $|m_3^2 - m_1^2|$, $\theta_{23}$,
$m_2^2 - m_1^2$, and $\theta_{12}$ have been measured,
and the upper bound on $\sin^2 2\theta_{13}$ has been obtained.
The goal of the next generation neutrino oscillation experiments
is hence to determine the remaining parameters, ${\it i.e.}$,
the sign of $m^2_3 - m^2_1$ (the neutrino mass hierarchy pattern),
a finite value of the third mixing angle $\theta_{13}$,
and the CP violating phase of the lepton sector, $\delta$.
\par
Several neutrino oscillation experiments 
are starting or being prepared
aiming at the measurements of $\theta_{13}$,
which can be categorized into two types;
one is the next generation of reactor neutrino experiments,
and the others are accelerator based neutrino oscillation experiments.
Three experiments,
Double CHOOZ \cite{double-chooz},
RENO \cite{reno}, and Daya Bay\cite{daya-bay},
will be able to observe the oscillation (disappearance)
of $\bar{\nu}_e$'s from reactors at about 1km distances
if $\sin^2 2\theta_{13} \simgt 0.01$.
In accelerator based LBL experiments, $\nu_{\mu}$ or
$\bar{\nu}_{\mu}$ beam
from high energy $\pi^{+}$ or $\pi^{-}$
decay-in-flight, respectively,
will be detected at a distance of a few to
several hundred km's away 
and 
the
$\nu_{\mu} \to \nu_e$ or $\bar{\nu}_{\mu} \to \bar{\nu}_e$ transition is
looked for.
The transition rate is proportional to $\sin^2 \theta_{13}$,
and the 
next generation LBL experiments,
the Tokai-to-Kamioka (T2K) experiment \cite{T2K}
and the NO$\nu$A experiment \cite{nova},
have a chance to discover the \mutoe~ transition,
if $\sin^2 2\theta_{13}$ \simgt 0.02\footnote{
Recently, the T2K experiment \cite{t2k-recent}
 and the MINOS experiment \cite{minos-recent} reported hints of
 non-zero $\theta_{13}$, based on observation of $\nu_\mu \to \nu_e$
 candidate events. 
}.
\par
\begin{figure}
\begin{center}
\includegraphics[angle = 0 , width=13cm]{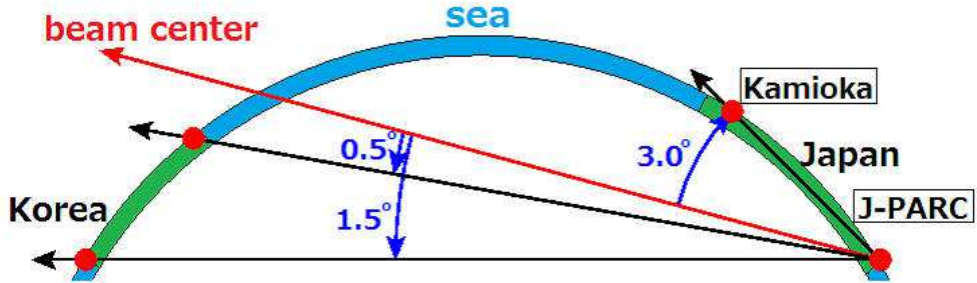}
\end{center}
\caption{
Schematic cross section view of the T2KK experiment
when the center of the neutrino beam from J-PARC
is 
\deg{3}
below the Super Kamiokande detector. 
The center of the beam is shown
by the red arrows,
and the black arrows show the baselines of a few off-axis beams (OAB).
}
\label{fig:new-cross}
\end{figure}

In this report,
we focus our attention on the proposed one-beam two-detector experiment,
the Tokai-to-Kamioka-and-Korea (T2KK) experiment
\cite{hagiwara-see-saw, t2kk, t2kr-l, t2kk-full, t2kk-atm, t2kk-bg}.
T2KK is an extension of the T2K experiment
where 
an additional huge detector is placed in Korea along the T2K
neutrino beam line between J-PARC at Tokai
and Super-Kamiokande (SK) at Kamioka.
We show the cross section view of the T2KK experiment in \figref{fig:new-cross}.
The center of the T2K neutrino beam has been designed to go through
the underground beneath SK
so that an off-axis beam (OAB) between~ \deg{2} and \deg{3} upward
from the beam center can be observed at SK,
in order to maximize the sensitivity to the 
\mutoe oscillation.
It then follows that the center of the T2K beam reaches the sea level
in the Japan/East sea,
and the lower-side of the same beam at \deg{0.5} or larger off-axis
angle goes through Korea \cite{hagiwara-see-saw, t2kr-l}.
Therefore, if we place an additional neutrino detector in Korea along the T2K beam line,
we can perform two LBL experiments at the same time.
It has been demonstrated that 
the T2KK experiment has an outstanding physics potential
to determine the neutrino mass hierarchy pattern,
normal ($m^2_3 - m^2_1 > 0$) or inverted ($m^2_3 - m^2_1 < 0$),
and to determine the CP violating phase $\delta$
\cite{t2kk, t2kr-l, t2kk-full}.
\par
The key of determining the neutrino mass hierarchy pattern
in the T2KK experiment
is the earth matter effect,
which arises from the coherent interaction
of $\nu_e$ (or $\bar \nu_e$) off the electrons in the earth matter \cite{matter-effect}.
The matter effect
enhances or suppresses the \mutoe~ oscillation probability, respectively,
if the mass hierarchy pattern is normal ($m^2_3 - m^2_1 > 0$) or inverted ($m^2_3 - m^2_1 < 0$).
Furthermore, the matter effect around the first oscillation maximum in Korea ($L \simgt 1000 $ km)
is significantly larger than the one at SK ($L = 295$ km)
because its magnitude grows with neutrino energy,
and the oscillation phase is roughly proportional to the ratio $L/E$.
Fortunately, we can observe the first oscillation maximum at both detectors
efficiently, if a detector is placed in the east coast of Korea such that
$0.5^{\circ} \sim 1^{\circ}$
OAB can be observed at $L \sim 1000$ km when the beam is adjusted to give the
$3.0^{\circ} \sim 2.5^{\circ}$  OAB at SK \cite{t2kr-l, t2kk-full}.
In a previous work \cite{t2kk-full},
it has been found that
the neutrino mass hierarchy can be determined most effectively
by observing $0.5^{\circ}$ off-axis beam in Korea
at $L \sim 1000$ km
and $3.0^{\circ}$ off-axis beam at Kamioka.
With this combination, the neutrino mass hierarchy
pattern can be determined at 3-$\sigma$ level
when $\sin^2 2\theta_{13} \simgt 0.06$,
if we place a 100 kt level Water \cerenkov detector in Korea
during the T2K experimental period with $5 \times 10^{21}$
POT (protons on target).
This is the minimum scenario where no additional detector is placed
in Kamioka and no increase is the J-PARC beam power is assumed.
A 100 kt level fiducial volume for the detector in Korea
is necessary in order to balance the statistics
of SK with 20 kt at 1/3 the distance from the J-PARC.
A more grandiose scenario with huge detector both
at Kamioka and Korea has been considered in Ref. \cite{t2kk}.
\par
Since the earth matter effect is very important in estimating the physics
discovery potential of the T2KK proposal,
we study in this report its impacts and uncertainty in detail
by using the recent geophysical measurements
of the earth matter density beneath Japan, Japan/East sea, and Korea.
Both the average and the distribution of the
matter density along the Tokai-to-Kamioka and the Tokai-to-Korea
baselines are studied, as well as their errors.
We also examine the impact of using both of the neutrino and anti-neutrino
beams,
since the earth matter effect on
the $\nu_{\mu} \to \nu_e$ and $\bar{\nu}_{\mu} \to \bar{\nu}_e$  
transitions have opposite signs.
\par
This paper is organized as follows.
In section 2, we study the earth matter density
profile along the Tokai-to-Kamioka and Tokai-to-Korea baselines,
and then we evaluate the average and the Fourier modes
of the matter densities along the baselines.
In section 3,
we discuss how to treat neutrino oscillations with non-uniform matter distributions.
We obtain both the exact solution to the oscillation amplitudes and
on approximate analytic expression for the oscillation probability that makes
use of the Fourier series of the matter distribution along the baseline.
In section 4, we introduce our analysis method
that quantifies the sensitivity of the T2KK experiment
on the neutrino mass hierarchy pattern by
introducing a $\chi^2$ function that accounts for both
statistical and systematic errors.
In section 5, we discuss the matter effect dependence
of the capability of determining the mass hierarchy pattern
in the T2KK experiment.
In section 6, we consider the matter effect for the anti-neutrino
experiment, and
examine the impact of
combining both neutrino and anti-neutrino oscillation experiments.
Finally, we summaries our findings in section 7.
In Appendix, we give the second-order perturbation formula for
the neutrino oscillation amplitudes
in terms of the matter effect term and the smaller mass squared difference.
\section{Earth matter profile along the T2KK baselines}
\hspace{11pt}
In this section, we study the matter density distribution along the baselines of Tokai-to-Kamioka
and Tokai-to-Korea, and then estimate the average and the Fourier modes of
the matter profile along the baselines.
\par
\subsection{Tokai-to-Kamioka baseline}
\par
\begin{figure}
\begin{center}
\includegraphics[angle = 0 , width=13cm]{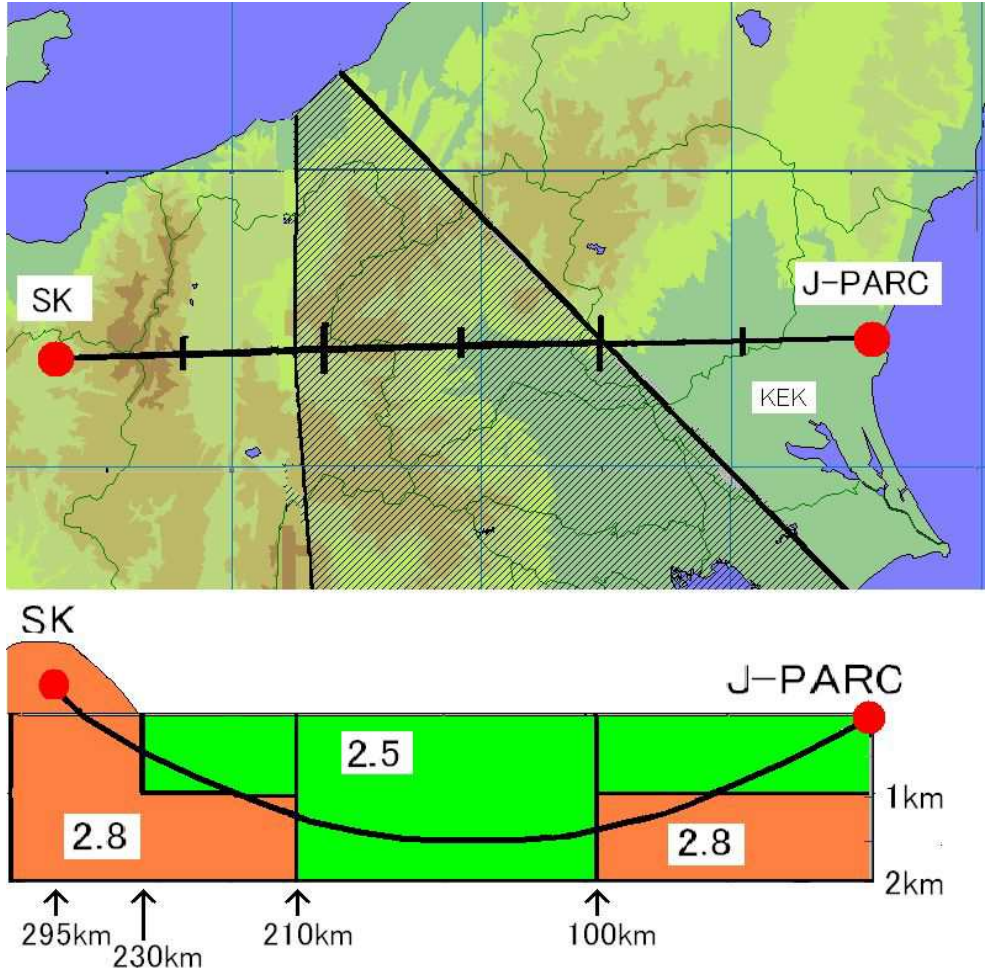}
\end{center}
\caption{
Upper: The Tokai-to-Kamioka (T2K) baseline.
The distance from J-PARC along the baseline is shown by tick marks every 50 km.
The shaded region is called Fossa Magna \cite{fossa-magna},
where the North American plate and the Eurasian plate
meet.
Lower:
The cross section view of the T2K experiment. The horizontal line shows the 
distance from the J-PARC \cite{j-parc} and the vertical axis measures the depth from the sea level.
The number in each region shows the mean matter 
density in units of $\density$.
The neutrino beam line between J-PARC and Kamioka is shown by an arc.
}
\label{fig:t2k-profile}
\end{figure}

\hspace{0.5cm}
First,
we show the Tokai-to-Kamioka (T2K)
baseline and
its cross section view
along the baseline
in \figref{fig:t2k-profile}.
In the upper figure, 
the scale along the baseline
shows the distance from J-PARC by tick marks every 50 km.
The shaded region is called Fossa Magna \cite{fossa-magna},
which is the border between the two continental plates, North American Plate and Eurasian Plate.
In this region, the sediment layer is estimated as 6 times deeper than
that in the surrounding region.
The lower plot shows the cross section view
of the T2K experiment. The horizontal line shows the 
distance from the J-PARC and the vertical axis gives the depth
from the sea level.
The number in each region stands for the average matter density in units of $\density$.
In this figure, the thickness of the sediment layer except for the Fossa Magna region
is assumed to be $1 \km$ \cite{koike98} and we refer to the geological map in Ref.~\cite{t2k-matter}.
In Fossa Magna, the sediment layer is as deep as $6$ km \cite{ fossa-magna},
which is mainly composed of limestone and sandstone with others \cite{t2k-matter}.
According to Ref.~\cite{fossa-magna}, the density of such sediment layer is $2.5 \density$.
On the other hand, the sediment layer near Kamioka, $L > 230$ km along the baseline,
is mainly composed
of granite
whose density is about $2.8 \density$.
\par
When we average out the matter density along the baseline
shown in \figref{fig:t2k-profile}, we obtain the average density of 2.6 $\density$,
which is significantly lower than the value 2.8$\density$ quoted
in the Letter of Intent (LOI) of the T2K experiment \cite{T2K}.
The difference is mainly due to Fossa Magna,
whose lower density has not been taken into account in the past.
\par 
The error of the average density
can be estimated from
the error of the mean density in each region,
and the uncertainty of the boundary of each layer.
In most of research works in geophysics, these two values are evaluated
by using the seismic wave observation.
The mean value of the matter density
is estimated from the velocity of the seismic wave
in each region.
In this work, 
we adopt the density-velocity correlation of Ref.~\cite{dv-conversion}
to estimate the matter density from the $p$-wave velocity.
\begin{figure}
\begin{center}
\includegraphics[angle = 0 , width=10cm]{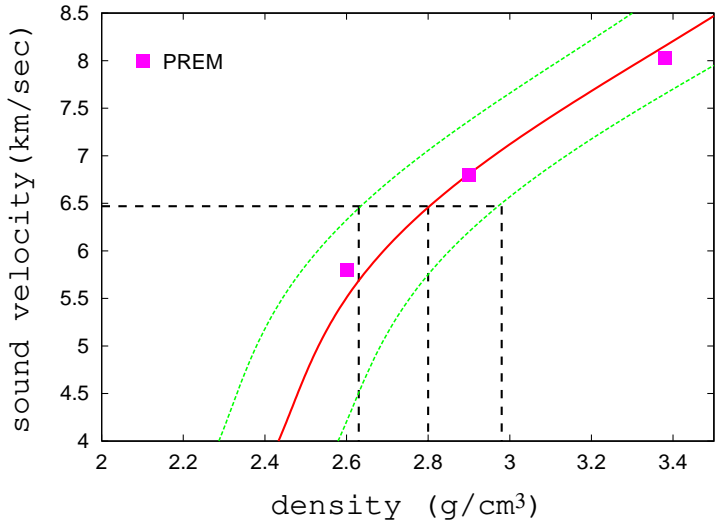}
\end{center}
\caption{
The relation between the matter density and the $p$-wave sound velocity
($V_p$) according to Ref.~\cite{dv-conversion}.
The solid line gives the mean value of the density, and the dotted lines in both sides
show the error of the estimated density.
The dashed lines show how we estimate the density and its error
when $V_p = 6.5$ km/sec.
The three square points are the reference points in Preliminarily
Reference Earth Model (PREM) \cite{prem}.
}
\label{fig:v-rho-trans}
\end{figure}
In \figref{fig:v-rho-trans},
we show the density velocity relation.
The solid line shows the relation between the sound velocity
and the mean matter density according to Ref.~\cite{dv-conversion},
which can be expressed as
\begin{equation}
\rho = -0.00283 V_p^4 + 0.0704 V_p^3 - 0.598 V_p^2 + 2.23 V_p - 0.7 \,,
\label{eq:v-rho-trans}
\end{equation}
where $V_p$ is the $p$-wave sound velocity inside the matter
in units of km/sec,
and the matter density $\rho$ is in units of $\density$.
The dotted lines in both sides show the error of the estimated density,
which is about $6\%$.
The dashed lines show how we estimate the density and its error
when $V_p = 6.5$ km/sec.
The three squares are the reference points shown in Preliminarily
Reference Earth Model (PREM) \cite{prem}, which is often used
to estimate the average density along the baseline for various LBL experiments.
These points lie close enough to the mean density line of \eqref{eq:v-rho-trans},
confirming the consistency between the PREM and the other geological measurements.
On the other hand,
the location of the boundary is evaluated from the reflection point
of the seismic wave.
The error of the boundary depth between the sediment layer and
the upper crust is estimated as $\pm 300 $ m.
It affects the average density along the T2K baseline
only by 
0.05$\%$,
and it can be safely neglected.
The error of the assumed sound velocity measurement at various location
can also be neglected.
In this report, we assume that the matter density error is determined
by the $6\%$ systematic uncertainty in the density-velocity relation,
which is taken to be common ($100\%$ correlation) at all location
along the baseline. On the other hand, the error of the average
matter density along the T2K baseline and the Tokai-to-Korea
baselines are taken to be independent in order to make
the most conservative estimate.
The $6\%$ error adapted in this report is a factor
of two larger than the error adapted in Refs.~\cite{t2kr-l, t2kk-full}. 
The matter density error is expected to become
smaller by making use of
the information from the
other measurements, {\it e.g.} the gravity anomaly,
the magnetic anomaly, and by actually digging into the crust.
\begin{figure}
\begin{center}
\includegraphics[angle = 0 ,width=10cm]{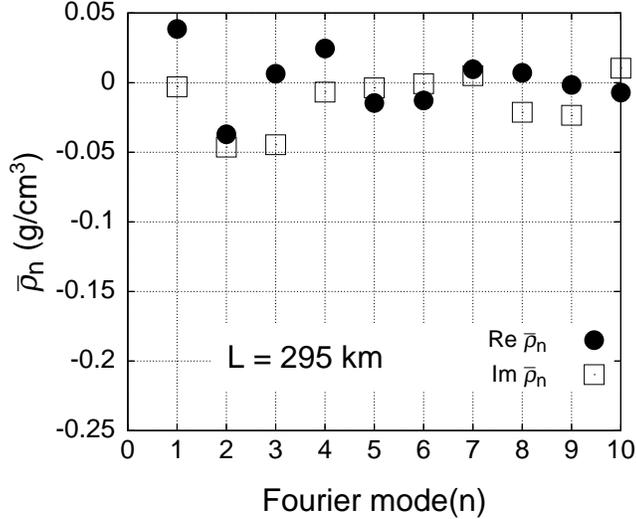}
\end{center}
\caption{
Fourier modes of the matter density distribution along the Tokai-to-Kamioka baseline
in units of $\density$.
Black circles and white squares show the
real and the imaginary parts of the Fourier coefficients, respectively.
}
\label{fig:fourie-295}
\end{figure}
\par
We note that
the density fluctuation along the T2K baseline of \figref{fig:t2k-profile} from the average
of 2.6 $\density$ is about $8\%$,
which is comparable to the $6\%$ error of the average matter density.
It is therefore of our concern that the
approximation of using only the average matter density along the
baseline may not be accurate enough.
In Ref.~\cite{koike98}, the authors have shown that the Fourier analysis
is useful to study the 
impacts of the matter density distribution.
We therefore show the Fourier coefficients of the matter density profile along the Tokai-to-Kamioka baseline
in \figref{fig:fourie-295}.
Black circles and white squares show the
real and the imaginary parts of the Fourier modes
in units of $\density$, respectively,
which are defined as
\begin{eqnarray}
\rho(x) &=& \sum_{k = -\infty}^{ \infty} \bar{\rho}_k {\rm exp}
\left(-i  \frac{2 \pi k x}{L} \right) \nonumber \\
&=& \bar{\rho}_0
+
2 \sum^{\infty}_{k = 1}
\left[ {\rm Re} 
\left(\bar{\rho}_k \right) 
\cos \left( \frac{2 \pi kx}{L} \right) 
+ {\rm Im} \left( \bar{\rho}_k  \right)
\sin \left( \frac{2 \pi kx}{L} \right)
\right]
\,,
\label{eq:fourie-rho}
\end{eqnarray}
where $\rho(x)$ is the matter density along the baseline
at a distance $x$ from J-PARC.
Positive value of Re($\bar{\rho}_1$) is clearly due to the
low density of Fossa Magna in the middle of the baseline,
whereas the negative values of Im($\bar{\rho}_2$) and Im($\bar{\rho}_3$)
reflect the high matter density in the Kamioka area that
makes the distribution asymmetric about the half point
$(x = L/2 )$ of the baseline.
We note, however, that
the magnitude of all the Fourier modes
is less than 0.05 $\density$,
which is less than $2\%$ of the average matter
density of $\bar{\rho}_0 = 2.6 \density$.
We can therefore expect that
the fluctuation effect can be
safely neglected in the T2K experiment,
as will be verified below.
\par
\subsection{Tokai-to-Korea baselines}
Let us now examine the matter density distribution
along the Tokai-to-Korea baseline.
\begin{figure}
\begin{center}
\includegraphics[angle = 0 ,width=15cm]{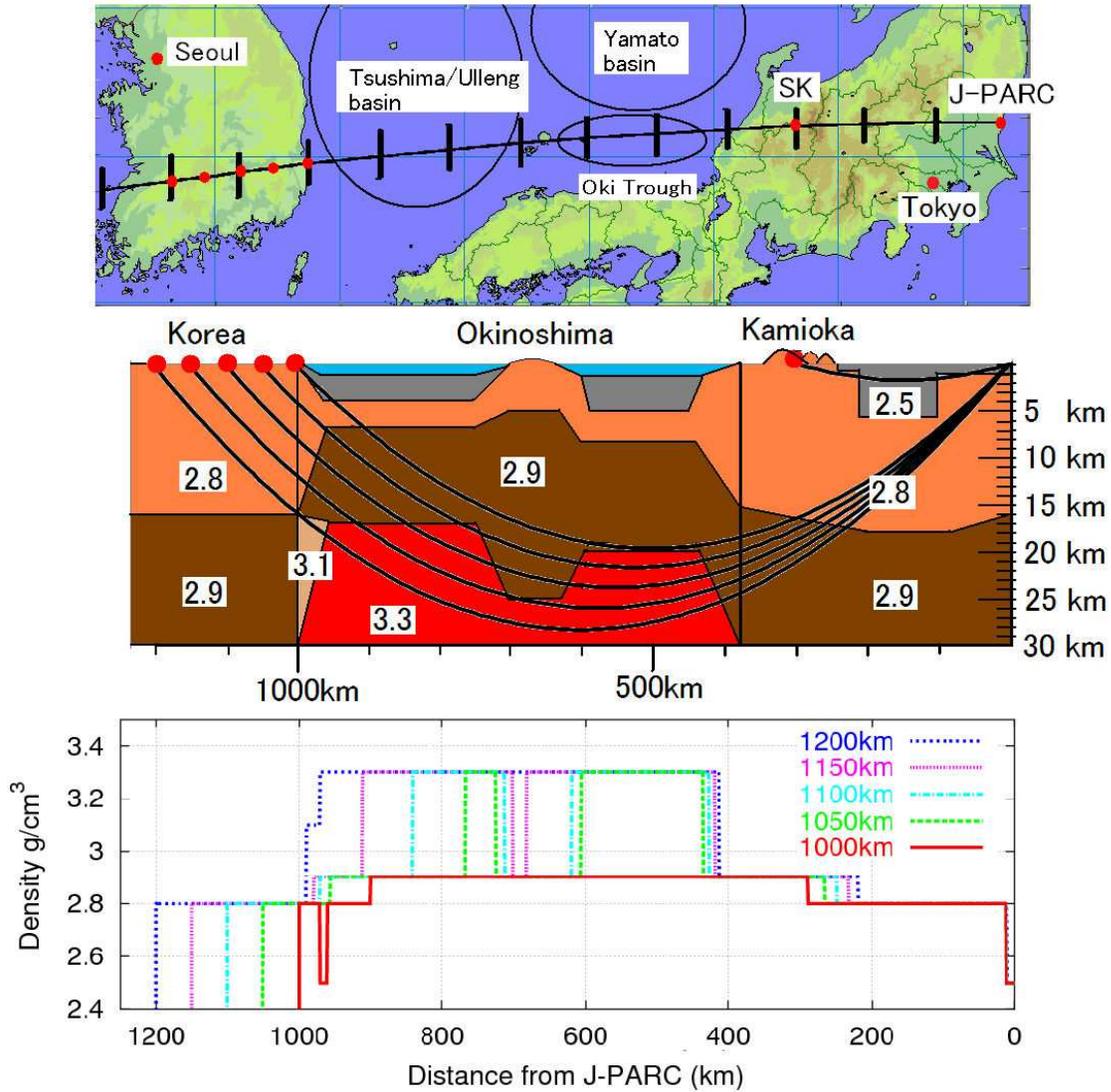}
\end{center}
\caption{
The surface map (up) of the Tokai-to-Kamioka-and-Korea (T2KK) experiment,
the cross section view (middle) and the density distribution
along the baselines (bottom).
The center of the neutrino beam line is shown in the upper figure,
along which the distance from J-PARC is given by the tick marks
every 100 km.
The three circles show the regions which have been studied by geophysicists;
Yamato basin, Oki trough \cite{yamato} and Tsushima/Ulleung basin 
\cite{oki, tsushima}.
In the middle figure,
the five curves show the baselines for $L = 1000$, 1050, 1100, 1150 and 1200 km,
where
the horizontal axis gives
the distance from J-PARC along the arc of the earth surface,
and the vertical scale measures the location of the baseline
below the sea level.
The numbers in white squares represent the average matter density
of the surrounding stratum, in units of
${\rm g}/ {\rm cm}^3$.
}
\label{fig:t2korea}
\end{figure}
In \figref{fig:t2korea},
we show the
surface map of the T2KK experiment (up),
its cross section view (middle),
and the matter density distribution 
along the baselines (bottom).
Along the center of the neutrino beam line shown in the surface map,
the distance 
from J-PARC is given by tick marks every
100 km.
The areas enclosed by 
the three circles show the regions which have been studied by geophysicists;
Yamato basin, Oki trough \cite{yamato} and Tsushima/Ulleung basin 
\cite{oki, tsushima}.
In the middle figure,
the five curves correspond to the baselines for $L = 1000$, 1050, 1100, 1150 and 1200 km,
where the horizontal axis measures
the distance along the earth arc from J-PARC,
and the vertical axis measures the location of the baselines below the sea level.
The numbers in white squares represent the average matter density, in units of
${\rm g}/ {\rm cm}^3$, of the surrounding stratum:
the sediment layer (2.5), the upper crust (2.8), the lower crust (2.9),
and the upper mantle (3.3).
A lump of 3.1 $\density$ at Korean coast is 
a lava boulder \cite{tsushima}.
In order to produce this figure, we refer to the recent study on the 
Conrad discontinuity, the boundary between the upper and the lower crust,
and on the \mohorovicic discontinuity (Moho-discontinuity), the boundary between the crust and the mantle, below Japan 
and Korea \cite{japan-profile, korea-profile}.
Regarding the matter profile beneath the Japan/East sea,
we divide the sea into three regions, Oki trough (east side) \cite{yamato},
Oki island (middle) \cite{oki}, and Tsushima/Ulleung basin (west side)\cite{tsushima}.
According to Refs.~\cite{yamato, tsushima}, typical depths
of the Conrad and the Moho-discontinuities
below Oki trough (Tsushima/Ulleung basin)
are 8 km (7km) and 20 km (17 km), respectively.
On the other hand,
Ref.~\cite{oki}
shows that they
are 6km and 25 km, respectively, below the sea level at around Oki island.
Although there is no direct measurement of the discontinuities along the
Tokai-to-Korea baselines,
we may assume that the depth of the discontinuities
is roughly characterized by the depth of the seabed of each region.
If the sea is shallower, the discontinuities should be deeper,
and {\it vice versa}.
Because the seabed above the baseline is rather flat the east and west
sides of Oki islands,
we obtain the estimates of \figref{fig:t2korea} from
the measured depths of the discontinuities in each region.
%
\par
In addition,
we adopt the following simple interpolations around
the sea bounds.
As for the Moho-discontinuity, we assume that
it is connected from 32 km below the sea level at the Japanese coast,
the point (380 km, 32 km) in the middle figure,
to 20 km below the Japan/East sea
at (440 km, 20 km)
by a straight line as illustrated in the figure.
It is at this point, about 60 km west from the coast along the beam direction,
the depth of the Japan/East sea reaches 1 km, the average depth of Oki trough.
As for the Oki island region,
the Moho-discontinuity is connected by straight line
from (600 km, 20 km) to (630 km, 25 km),
flat from 630 km to 700 km,
and then by another straight line from (700km, 25km) to (730km, 17 km).
Near the Korean coast,
it is connected by a straight line from (970km, 17km)
to (990km, 32 km).
Here
the slope is shaper because the Korean coast
is only about 20 km away from the western edge of Tsushima/Ulleung basin \cite{tsushima}.
As for the Conrad discontinuity,
we also assume the linear interpolation between
(380 km, 15 km) and (440 km, 8 km) at the Japanese coast,
(600km, 8km) and (630km, 5km) at east side of Oki island region,
(700km, 5km) and (750km, 7km) at west side of Oki island region,
and between (970 km, 7 km) and (990 km, 17 km) at the
Korean coast.
The lava bounder lies on top of the oblique section
of the Moho-discontinuity and it is assumed
to have the triangle shape along the cross section
with the apexes at, (970 km, 17 km), (990km, 15km), and (990km, 30km)
as also illustrated in the middle figure.
Although the above treatments of the boundary conditions are very rough,
we do not elaborate them further because the neutrino oscillation probabilities are very
insensitive to the exact location of the boundaries.
\par
With the above setting, we obtain the matter density distribution
along the baselines shown in the bottom figure of \figref{fig:t2korea}.
Here the horizontal axis measures the distance $x$ from J-PARC along the baseline.
We assume that a detector in Korea is placed at the sea level
(zero altitude) along the plane of the center and the center of the earth,
at the distance $L$ from J-PARC.
Along the baseline of length $L$,
the point $x$ (km) away from J-PARC appears on the cross section figure
in the middle of \figref{fig:t2korea} at the distance $l$ (km) along the earth arc,
and the depth $d$ (km) below the sea level;
\begin{subequations}
\label{eq:position}
\begin{eqnarray}
l &=& R\arcsin \left(\frac{x}{R} \sqrt{
\displaystyle\frac{R^2 - L^2/4}{R^2-xL+x^2}} \right)
\nonumber \\
&=&x \left\{1  
- \frac{x^2}{3R^2} + \frac{xL} {2R^2} - \frac{L^2}{8R^2}
+ {\it O}\left( R^{-4} \right)
\right\} ,
\label{eq:distance}
\\
d &=& R - \sqrt{ x^2 - x L + R^2 }   =  \frac{x\left(L - x \right)}{2R}
        \left[ 1 + {\it O}\left( R^{-3} \right)  \right]\,,
\label{eq:depth}
\end{eqnarray}
\end{subequations}
where $R$ is the radius of the earth, 6378.1 km.
The five arcs are the cross section view of
\figref{fig:t2korea} are then obtained parametrically
as ($l(x), d(x)$).
It is also noted that $l$ is not much different 
from $x$ for $L \sim 1000$ km;
$l$ is about $0.01 \%$ shorter
than $x$ at $\sim 300$ km, and $0.1 \%$ longer $x$ at $x \sim 1000$ km.
The effect of the lava boulder near the Korean coast is seen in the
matter density distribution for $L=$ 1100 km and 1150 km.
We also notice that the baseline for $L = 1000$ km goes through the sediment layer
(2.5 $\density$) near the Korean east at $970 {\rm km} \simlt l \simlt 990$km
($969 {\rm km} \simlt x \simlt 989 {\rm km}$).
All the baselines go through the sediment layer near the J-PARC
until the baseline reaches the depth of $d =$ 1 km
($0 \simlt x \simlt 10 \sim 13$ km).
\par
Let us now estimate the average matter density and its error along each baseline.
Since the traveling distance in the mantle and that in the crust depend on
the locations of their boundaries,
the error of the average matter density is affected by the errors of the depths of
the discontinuities.
It is especially important for
the Moho-discontinuity
because
the density difference between the lower crust and the upper mantle
is as large as $0.4 \density$,
which is about $13\%$ of their average.
Therefore, 
we examine the impacts of varying
the depth of the Moho-discontinuity
on the average matter density.
\begin{table}[t]
\begin{center}
\begin {tabular}{|c||c|c|c|c|c|} \hline
Baseline length & \multicolumn{5}{c|}{The depth of the Moho-discontinuity below the sea level} \\ \hline
$L$ & $-3 \sigma$ & $-1\sigma$  &  mean  & $+1\sigma$ & $+3 \sigma$  \\ \hline
1000 km    &~~2.923 ~~&~~2.900 ~~&~~2.854 ~~&~~2.854 ~~&~~2.854 ~~\\ \hline 
1050 km    & 2.973 & 2.954 & 2.944 & 2.933 & 2.862 \\ \hline 
1100 km    & 3.001 & 2.988 & 2.980 & 2.971 & 2.952 \\ \hline 
1150 km    & 3.049 & 3.039 & 3.027 & 3.002 & 2.986 \\ \hline 
1200 km    & 3.054 & 3.052 & 3.051 & 3.049 & 3.036 \\ \hline
Re$\bar{\rho}_1$ $ (L = 1000$ km) & $-0.097$ & $-0.076$ & $-0.031$ & $-0.031$ & $-0.031$
\\ \hline
Re$\bar{\rho}_1$ $ (L = 1050$ km) & $-0.102$ & $-0.098$ & $-0.094$ & $-0.090$ & $-0.028$
\\ \hline
Re$\bar{\rho}_1$ $ (L = 1100$ km) & $-0.117$ & $-0.110$ & $-0.101$ & $-0.105$ & $-0.099$
\\ \hline
\end{tabular}
\end{center}
\caption{Average matter density
along the baselines
of $L = 1000$, 1050, 1100, 1150 and 1200 km,
in units of $\density$,
for five possible
locations of the Moho-discontinuity, which is the boundary
between the lower crust and the upper mantle.
The bottom three rows give the real part of the first Fourier coefficient of
the matter density distribution along the $L = 1000, 1050,$ and 1100 km baselines.
}
\label{tab:average}
\end{table}
In \tableref{tab:average},
we show
the
average matter density along the baselines
of
$L = 1000$, 1050, 1100, 1150 and 1200 km,
in units of $\density$,
for five
locations of the Moho-discontinuity.
The mean corresponds the depth of 20 km at $L \sim 500$ km
and 17 km at $L \sim 800$ km in \figref{fig:t2korea},
$\pm 1 \sigma~(\pm 3 \sigma)$ corresponds to the overall
shifts by $\pm 0.7$ km ($\pm2$ km) of the Moho-discontinuity depth
below the sea level.
\par
Because longer baselines go through the upper mantle for a longer distance,
the average density grows with $L$;
2.854 $\density$ at  $L = 1000$ km to
2.944, 
2.980, 
3.027, 
and 3.051 $\density$ at $L (\rm km)=$
1050, 1100, 1150, and 1200,
respectively.
The average density at $L = 1000$ km
is quite sensitive to the rise of the Moho-discontinuity
because the baseline almost touches the  mantle
at $L \sim 500$ km
for the mean estimate, as shown in \figref{fig:t2korea}.
The average density grows from $2.854 {\rm g/cm}^3$
for the mean depth
to $2.900 {\rm g/cm}^3$, by $1.6\%$,
if the Moho-discontinuity is only 0.7 km ($1 \sigma$)
higher than the present estimate.
Although this is striking, the effect is significantly
smaller than the $6\%$ overall uncertainty
in the conversion of the sound velocity to the matter density.
We therefore neglect the effect in the present study,
but further geophysical studies may be desired if $L \sim 1000$ km
baseline is chosen for the far detector location.
\par
We also examine the impacts of 
the non-uniformity of the matter density distribution along the Tokai-to-Korea baseline.
\begin{figure}
\begin{center}
\includegraphics[angle = 0 , width = 7.8cm]{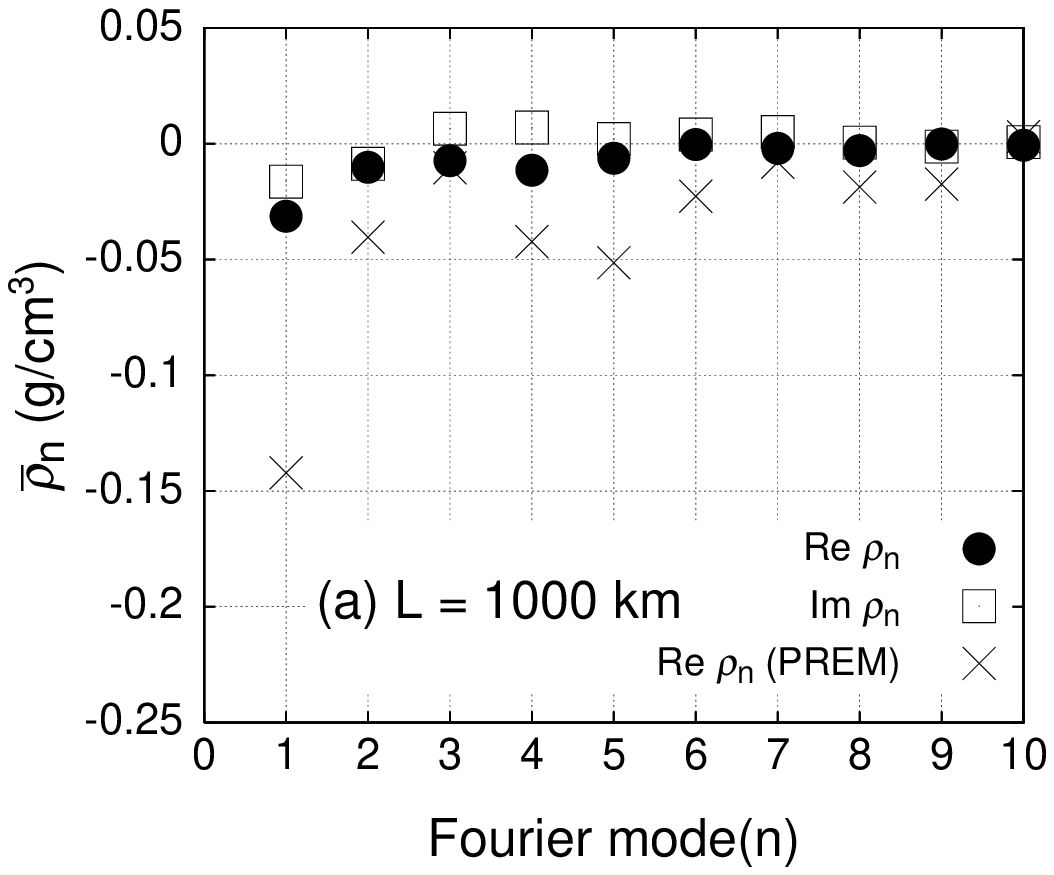}
\includegraphics[angle = 0 , width=  7.8cm]{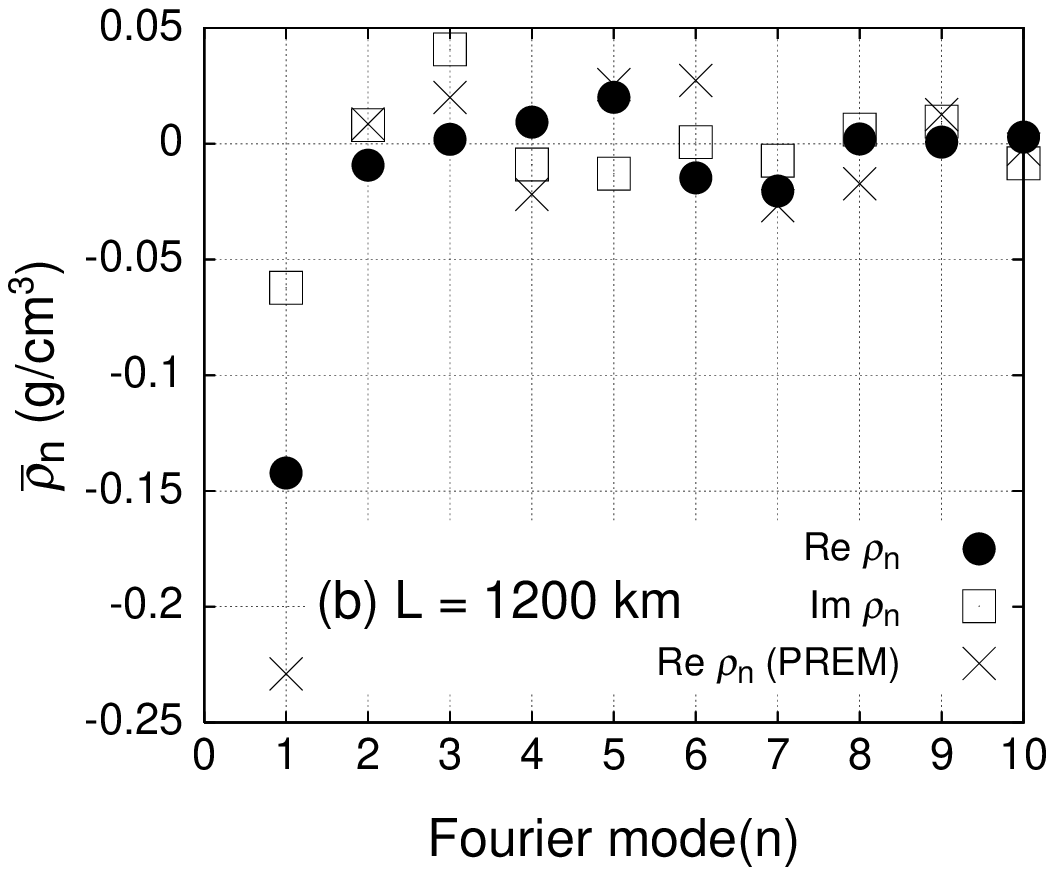}
\end{center}
\caption{
Fourier coefficients of the density distribution along the Tokai-to-Korea
baselines ;~$L = 1000$ km (a), and 1200 km (b).
Black circles and white squares show the
real and imaginary parts of the Fourier coefficients, respectively,
and the cross symbols show the real parts estimated from PREM \cite{prem}. 
}
\label{fig:f-real}
\end{figure}
In \figref{fig:f-real},
the black circles and white squares show,
respectively, the real and imaginary parts of
the Fourier coefficients of the matter density distribution along the
baselines of $L = 1000$ km (a) and 1200km (b).
As a reference, we show by the cross symbols
the real part of the Fourier coefficients of the
matter density distribution of PREM \cite{prem}.
The PREM generally gives symmetric matter distribution
about $x = L/2$, 
and hence does not give imaginary cofficients.
When we compare them with the distribution for the Tokia-to-Kamioka baseline
of \figref{fig:fourie-295},
we find that
Re$(\bar{\rho}_1)$ is now negative
and its magnitude grows from 0.03 $\density$
at $L = 1000$ km
to 0.14 $\density$ at $L = 1200$ km.
We note here that
Re($\bar{\rho}_1$)
at $L = 1000$ km depends strongly on
the depth of the Moho-discontinuity,
because whether the baseline goes through mantle or not
affects the convexity of the matter density distribution significantly;
see \figref{fig:t2korea}
and discussions on the average matter density above.
In order to show this sensitivity, we give Re$(\bar{\rho}_1$) values for
$L = 1000$, 1050, and 1100 km in the bottom three rows of \tableref{tab:average}.
It changes from $-0.03{\rm g/cm}^3$ to $-0.08 {\rm g/cm}^3$ at $L = 1000$ km
by more than a factor of two, if the Moho-discontinuity is only $0.7$ km ($1\sigma$)
shallower than our estimate of 20 km
based on the measurement in Oki trough \cite{yamato}.
As for the $L = 1050$ km, the Re($\bar{\rho}_1$) is almost insensitive to
the depth of the Moho-discontinuity. 
Since the baseline of $L = 1100$ km is more than 22 km below at Oki trough,
the magnitude of Re($\bar{\rho}_1$) of  $L = 1100$ km
is insensitive to the exact depth of the Moho-discontinuity.
\par
Along the Tokai-to-Korea baseline,
the convexity measure of $-{\rm Re} (\bar{\rho}_1)/\bar{\rho}_0$ grows
from $\sim 1\%$ ($\sim 3\%$ if the Moho-discontinuity is $3\%$ shallower
than our estimate) at $L = 1000$ km
to $5\%$ at $L = 1200$ km. Their effects are hence non-negligible even with
the $6 \%$ overall
uncertainty in the average matter density.
We also note that
the negative sign of Im$(\bar{\rho}_1)$
reflects the fact that
the mantle is slightly closer to a far detector in Korea
than J-PARC in Tokai.
\par
\subsection{Comparison with PREM}
Before closing this section, we compare the matter density distribution
of this work and those of the PREM \cite{prem},
where the earth is a spherically symmetric ball
such that the depth of the boundaries between adjacent layers
are the same everywhere.
\begin{figure}[t]
\begin{center}
\includegraphics[angle = 0 ,width = 14cm]{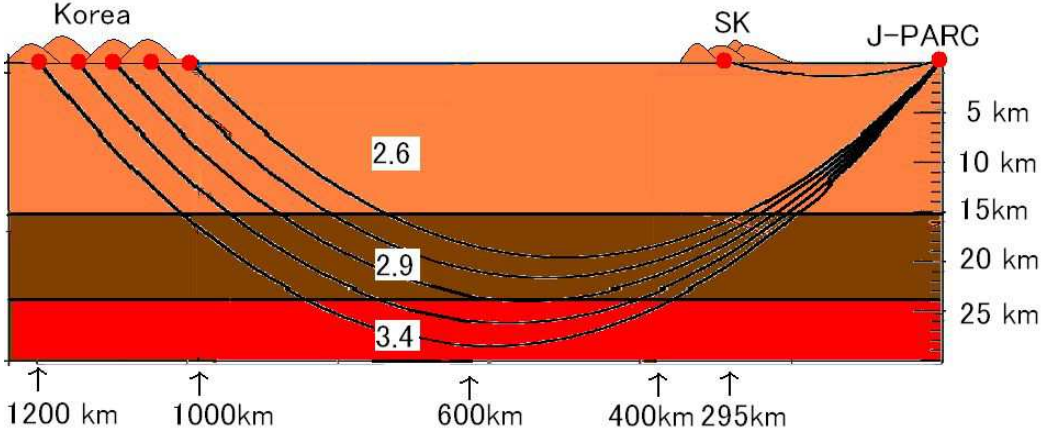}
\end{center}
\caption{
The cross section view of the T2KK experiment along the baselines 
according to Preliminarily Reference Earth Model (PREM) \cite{prem}.
}
\label{fig:prem}
\end{figure}
In \figref{fig:prem},
we show the cross section view of the T2KK experiment along the baselines
according to PREM.
Although the original PREM assumes that the sea covers whole of the earth down
to 3 km from the sea level, we assume in this figure that
none of the baselines go through the sea.
\par
There are several notable differences
between the matter density distribution of 
this work shown in \figref{fig:t2korea} and
that of PREM in \figref{fig:prem}.
As for the T2K baseline,
the Fossa Magna lies in the middle in \figref{fig:t2k-profile},
while it goes through only the upper crust in PREM.
More importantly,
the Conrad and the Moho-discontinuities
are significantly shallower than those of 
PREM below Japan/East sea,
whereas they 
are deeper than PREM below Japanese islands and Korean peninsula.
Accordingly, although all
but $L = 1000$ km
Tokai-to-Korea
baselines in \figref{fig:t2korea} 
go through the upper mantle,
the baselines of $L \simlt 1100$ km are
contained in the crust according to PREM.
\par
\begin{table}
\begin{center}
\begin {tabular}{|c|c|c|c|c|c|c|c|c|}
 \hline
    &\multicolumn{4}{c|}{This work} &  \multicolumn{4}{c|}{PREM} \\ \hline
$L$ (km) &  $\bar{\rho}_0$ &  ${\rm Re} \bar{\rho}_1$ & ${\rm Im} \bar{\rho}_1$ & ${\rm Re} \bar{\rho}_2$   
  &  $\bar{\rho}_0$ &  ${\rm Re} \bar{\rho}_1$ & ${\rm Im} \bar{\rho}_1$ & ${\rm Re} \bar{\rho}_2$ \\ \hline
295& 2.62 & $+0.039$ & $-0.003$ & $-0.037$ &
          2.60 & 0        & 0        & 0        \\ \hline
1000& 2.85 & $-0.031$ & $-0.016$ & $-0.010$ &
          2.71 & $-0.098$ & 0        & $-0.040$ \\ \hline
1050& 2.94 & $-0.094$ & $-0.028$ & $+0.030$ &
          2.76 & $-0.094$ & 0        & $-0.016$ \\ \hline
1100& 2.98 & $-0.108$ & $-0.044$ & $+0.003$ & 
          2.78 & $-0.090$ & 0        & $-0.030$ \\ \hline 
1150& 3.03 & $-0.134$ & $-0.060$ & $-0.001$ & 
          2.93 & $-0.201$ & 0        & $+0.040$ \\ \hline
1200& 3.05 & $-0.142$ & $-0.062$ & $-0.002$ &
          3.00 & $-0.220$ & 0        & $+0.009$ \\ \hline
\end{tabular}
\end{center}
\caption{Average and the Fourier coefficients of the matter density 
distribution along the Tokai-to-Kamioka baseline of \figref{fig:t2k-profile}
and the Tokai-to-Korea baselines in \figref{fig:t2korea},
which are compared with the predictions of PREM \cite{prem}.
All the numbers are in units of $\density$.
}
\label{tab:compare-prem}
\end{table}
These differences are reflected in the average
densities and in the  Fourier coefficients
of the matter distributions
along the baselines.
In \tableref{tab:compare-prem}, we summarize the average matter density
($\bar{\rho}_0$)
and a few Fourier coefficients along each baseline
estimated from the matter density profile of this work,
and those according to PREM, in units of $\density$.
\par
For the Tokai-to-Kamioka baseline, $L = 295$~km,
despite the difference in the distribution,
the average matter density of $2.62 \density$
turns out to be very similar to the upper crust density
of 2.6 $\density$ in PREM.
This is because the matter density around
Kamioka, being in the mountain area,
is higher than the PREM upper crust density,
which compensates for the low density in Fossa Magna.
The presence of Fossa Magna around the middle of
the baseline results in
the positive value of Re$(\bar{\rho}_1)$,
reflecting the concave nature of the Tokai-to-Kamioka
matter distribution.
\par
Along the Tokai-to-Korea baselines,
all the average densities estimated from \figref{fig:t2korea} are larger
than those of PREM, because
the crust density of $2.8 {\rm g/cm}^3$ in the region is
higher than the world average of $2.6 {\rm g/cm}^3$ in PREM,
and also because
the traveling length in the upper mantle
is longer than PREM, reflecting the shallower 
discontinuities below Japan/East sea in \figref{fig:t2korea}.
Especially for $L \sim 1050$ km, 
the average matter density of this work
is larger than PREM by $6.1\%$.
The difference reduces to $1.6\%$ at $L \sim 1200$ km.
The negative values of Re$(\bar{\rho}_1)/\bar{\rho}_0$
represent the convexity of the 
matter density distribution.
Their magnitudes are $\sim 1\%$ 
($\sim 3\%$ for $3\%$ shallower Moho-discontinuity)
at $L = 1000$ km and
$\sim 5\%$ at $L = 1200$ km
in this work, which are to be compared with 
$\sim 4\%$ and $\sim 7\%$,
respectively, according to PREM.
\section{Neutrino oscillation in the matter}
In this section, we explain how we calculate the neutrino
flavor oscillation probabilities with non-uniform matter distribution
efficiently, and introduce analytic approximations that help our understandings
on the matter effect.
\par
\subsection{Exact evaluation of oscillation probabilities}
First, we show how we compute the oscillation probability exactly
when the matter distribution is approximated by step functions.
In the earth, neutrinos interact coherently with electrons and nucleons
via charged and neutral weak boson exchange,
and these coherent interactions give rise to an additional potentials
in the Hamiltonian.
The potential from the neutral current interactions
are flavor-blind,
and it does not affect the neutrino flavor oscillation
probabilities.
Only the electron neutrinos,
$\nu_e$ and $\bar{\nu}_e$,
have coherent charged-current interactions with the electrons
in the matter.
This additional potential contributes only to 
the $\nu_e$ and $\bar{\nu}_e$ phase,
and hence affects the neutrino flavor oscillation probabilities
\cite{matter-effect}.
\par
The location ($x$) dependent effective Hamiltonian of a neutrino propagating
in the matter can hence be expressed as
\begin{eqnarray}
\label{eq:hamiltonian}
H (x) &=&
\frac{1}{2E}
\left[
U 
\left(
\begin{array}{ccc}
0 &  & \\
  & m^2_2 - m^2_1 & \\
  &   & m^2_3 - m^2_1
\end{array}
\right)
U^{\dagger} 
+ \left(
 \begin{array}{ccc}
a(x) &   &  \\
  & 0 &  \\
  &   & 0
\end{array}
 \right) 
 \right]
 \,,
\end{eqnarray}
on the flavor space
$\left(\nu_e,~\nu_\mu,~\nu_\tau \right)^{\rm T}$,
after removing the components which are
common to all the neutrino flavors.
Here $U$ is the lepton-flavor mixing matrix, the MNS matrix \cite{mns}, defined as,
\begin{equation}
\left| \nu_{\alpha} \right\rangle =  U_{\alpha i} \left|\nu{_i} \right\rangle \,,
\label{eq:state}
\end{equation}
where $\nu_{\alpha}$ represents the flavor eigenstate ($\nu_e, \nu_{\mu}$ and $\nu_{\tau}$) and
$\nu_i$ represents the mass eigenstates with the mass $m_i~(i = 1, 2, 3)$.
The term $a(x)$ in \eqref{eq:hamiltonian} represents the matter effect;
\begin{equation}
a(x) = 2\sqrt{2} G_F E n_e(x) \approx 7.56 \times 10^{-5} {\rm eV}^2 
\left(   \frac{\rho(x)}{{\rm g / cm}^3} \right)
\left(   \frac{E}{{\rm GeV}} \right)
\,,
\label{eq:mattereffect}
\end{equation}
where $G_F$ is the Fermi coupling constant,
$E$ denotes the neutrino energy,
and
$n_e(x)$ is the electron number density in the matter
at a location $x$ which is proportional to
the matter density in the earth, $\rho(x)$.
\par
Because of the location dependence of the matter density $\rho$,
the Hamiltonian \eqref{eq:hamiltonian} has explicit
dependence on the location $(x)$ of the neutrino
along the baseline,
or on the elapsed ($t=x/c$) since
the neutrino leaves J-PARC.
This time dependence can be easily solved
for the step-function approximation of the matter profile,
since the Hamiltonian is time-independent within
each layer of the region with a constant density\cite{ohlson}.
The time evolution
of the neutrino state can hence be obtained by
the time-ordered product of the evaluations
with time-independent Hamiltonian in each region;
\begin{equation}
e^{-i \int^L_0 H dx} = T \prod^n_{k=1}  e^{-i H_k (x_k - x_{k-1})} \,.
\label{eq:multi-time-evo}
\end{equation}
Here $T$ denotes the time ordering, $H_k$ is the time independent Hamiltonian
in the region $x_{k-1} \le x < x_{k}$ ($k = 1$ to $n$) along the baseline,
$x_{k}$ is the starting location of the region,
where $x_0= 0$ (km) is the location of the neutrino production point
at J-PARC and $x_n = L$ (km) is the location
of the target, SK or a far detector in Korea. 
\par
Although it is straightforward to
diagonalize the Hamiltonian
\eqref{eq:hamiltonian} in the region
$x_{k-1} \le x < x_{k}$ where the matter effect term
$a(x)$ is a constant,
we find that an efficient numerical code
can be developed by adopting a specific parametrization of the
MNS matrix which allows us to decouple the dependence of the
oscillation amplitude on the mixing angles $\theta_{23}$ and
the phase $\delta$ from the matter effects \cite{hagiwara-okamura}.
The standard parametrization of the MNS matrix can be expressed as
\begin{eqnarray}
&&\hspace{-2.75cm}U = O_{23}~P~O_{13}~P^{\dagger}~O_{12} \nonumber \\
  & \hspace{-4.0cm}=& \hspace{-2cm}
\left(
\begin{array}{ccc}
 1 &         &     \\
   & c_{23}  &s_{23} \\
   & -s_{23} &c_{23}
\end{array}
\right)
P
\left( \begin{array}{ccc}
c_{13} &        & s_{13}   \\
  & 1 &  \\
-s_{13} & & c_{13}
\end{array} \right)
P^{\dagger}
\left( \begin{array}{ccc}
c_{12} & s_{12}       &  \\
-s_{12}  & c_{12} &  \\
 & & 1
\end{array} \right)\,,
\label{eq:MNS-standard-para}
\end{eqnarray}
by using a diagonal complex phase matrix $P =$ diag $(1,~1,~ e^{i\delta})$,
where
$s_{ij}$ and $c_{ij}$ represent $\sin \theta_{ij}$ and $\cos \theta_{ij}$,
respectively.
The Hamiltonian $H_k$ of \eqref{eq:hamiltonian} with $a(x) = a_k$
can be
partially diagonalized
as
\begin{eqnarray}
H_k &=&
\displaystyle\frac{1}{2E}
O_{23}P
\left[
O_{13}O_{12}
\left(
\begin{array}{ccc}
0 &  & \\
  & m^2_2 - m^2_1 & \\
  &               & m^2_3 - m^2_1
\end{array}
\right)
O_{12}^{\rm T}O_{13}^{\rm T}
+
\left(
\begin{array}{ccc}
a_k & & \\
  & 0& \\
  & & 0
\end{array}
\right)
\right]
P^{\dagger} O_{23}^{\rm T}
\nonumber \\
&=&
\displaystyle\frac{1}{2E}
O_{23}P
\left[
\tilde{O}_k
\left(
\begin{array}{ccc}
0 & & \\
  & (\lambda_2 - \lambda_1)_k & \\
  & & (\lambda_3 - \lambda_1)_k
\end{array}
\right)
\tilde{O}^T_k
\right]
P^{\dagger} O_{23}^{\rm T}
+ \frac{\lambda_{1k}}{2E}
 \,,
\label{eq:rewrite-h}
\end{eqnarray}
where
$\lambda_{i k}$ ($i = 1,~2,~3$) are the eigenvalues
of the matrix inside the big parenthesis
and $\tilde{O}_k$ is 
the ordinal diagonalizing matrix.
Remarkable point 
of this parameterization
is that the parameters 
$\theta_{23}$ and $\delta$ are
manifestly independent of
the matter effect \cite{hagiwara-okamura}.
The decomposition \eqref{eq:rewrite-h}
allows us to compose a fast
and accurate numerical program for neutrino oscillation
in the matter,
since the non-trivial part of the Hamiltonian can be diagonalized by
using only real numbers.
\par
For the step-function like matter profile that gives \eqref{eq:multi-time-evo},
the oscillation amplitude can be solved explicitly as
\begin{eqnarray}
S_{\beta \alpha}(L)&=&
\left \langle \nu_{\beta} \right |
Te^{-i\int^{L}_{0} H dx}
\left| \nu_{\alpha} \right \rangle
\nonumber \\
&=&
\left \langle \nu_{\beta} \right |
O_{23}PT
\prod^{n}_{k= 1} \left[
\tilde {O}_k e^{-i \Delta \lambda_k \left(x_k - x_{k-1} \right)} \tilde{O}^{T}_{k}
\right]
P^{\dagger}O_{23}^{\rm T} 
\left| \nu_{\alpha} \right \rangle 
\times {\rm (phase)}
\,.
\label{eq:step-exact-s}
\end{eqnarray}
Here 
\begin{eqnarray}
\Delta \lambda_k
&=&
\displaystyle\frac{1}{2E}
\left(
\begin{array}{ccc}
0 & &                      \\
  &(\lambda_2 - \lambda_1)_k & \\ 
  & & (\lambda_3 - \lambda_1)_k
\end{array}
\right)
\,
\label{eq:new-eigen-vector}
\end{eqnarray}
gives the flavor non-universal part of the diagonal Hamiltonian
in the region $x_{k-1} \le x < x_{k}$ where the matter density $\rho$
and hence the matter effect term can be regarded as a constant
$a = a_k$.
Finally, we obtain the exact formula for the flavor transition probabilities,
\begin{equation}
P_{\nu_{\alpha} \rightarrow \nu_\beta}(L)
= \left|S_{\beta \alpha} (L) \right|^2 \,.
\label{eq:exact-p}
\end{equation}
It is explicitly seen in \eqref{eq:step-exact-s} that the dependence
on $\theta_{23}$ and $\delta$ of the oscillation amplitude and the transition
and survival probabilities are completely independent of the matter effect\cite{hagiwara-okamura}.
\subsection{Perturbation formula}
The exact solution of the time evolution operator, \eqref{eq:exact-p},  
is useful for numerical computation but
it does not illuminate our
understanding of the effect of
the matter density distribution.
In this subsection, we present
a perturbation formula for
the neutrino oscillation probability with non-uniform matter profile.
It helps us to understand not only the
mechanism of determining the neutrino mass hierarchy pattern
by using the average matter effect along the baselines,
but also the effects of the matter
density distribution along the baselines.
\par
Present experimental results tell that,
among the three 
terms of (mass)$^2$ dimension
in the Hamiltonian (\ref{eq:hamiltonian}),
$|m^2_3 - m^2_1| \sim 10^{-3} {\rm eV}^2$
is much larger than the other two terms,
$ m^2_2 - m^2_1 \sim 10^{-4} {\rm eV}^2$ and
$a \sim 10^{-4} {\rm eV}^2$ for $\rho \sim 3 \density$ and $E \sim 1$ GeV.
This suggests splitting of the Hamiltonian (\ref{eq:hamiltonian})
into the main term $H_0$ and the small term $H_1$:
\begin{subequations}
\label{eq:hamiltonian-all} 
\begin{eqnarray}
\label{eq:split-hamiltonian}
H(x) &=& H_0 + H_1(x)  \,, \\
\label{eq:hamiltonian0}
H_0 &=& \frac{1}{2E} U 
\left(
\begin{array}{ccc}
0 &   &  \\
  & 0 &  \\
  &   & m^2_3 - m^2_1
\end{array}
\right)
U^{\dagger} \,,
\\
\label{eq:hamiltonian1}
H_1(x) &=& \frac{1}{2E}
\left[
U
\left(
\begin{array}{ccc}
0 &   &  \\
  & m^2_2 - m^2_1 & \\
  &   & 0
\end{array}
\right)U^{\dagger}
 + \left(
 \begin{array}{ccc}
a(x) &   &  \\
  & 0 &  \\
  &   & 0
\end{array}
 \right) 
 \right]
 \,.
\end{eqnarray}
\end{subequations}
When $H_0 L$ is order unity, we can treat $H_1(x)$ as a small perturbation.
We examine the second order approximation:
\begin{eqnarray}
Te^{-i \int^{L}_0 H dx} &\approx&
e^{-iH_0 L}
- i  \int^{L}_0 dx e^{iH_0 (x - L)} H_1(x) e^{- i H_0x}
\nonumber
\\
&&-\int^{L}_{0} dx \int^{x}_{0} dy
e^{iH_0 (x-L)} H_1(x) e^{i H_0 (y-x)}H_1(y) e^{- i H_0y}
\,,
\label{eq:t-evolution}
\end{eqnarray}
where $H_0$ is location independent, and the location dependence of $H_1(x)$ is due to the matter profile $\rho(x)$,
or $a(x)$ along the baseline.
\par
The integral can be performed analytically by using the Fourier expansion of the
matter profile along the baseline \cite{koike98};
\begin{eqnarray}
a(x) &=& \sum_{k = -\infty}^{ \infty} \bar{a}_k {\rm exp}
\left(-i  \frac{2 \pi k x}{L} \right) \nonumber \\
&=& \bar{a}_0
+
2 \sum^{\infty}_{k = 1}
\left[ {\rm Re} 
\left(\bar{a}_k \right) 
\cos \left( \frac{2 \pi kx}{L} \right) 
+ {\rm Im} \left( \bar{a}_k  \right)
\sin \left( \frac{2 \pi kx}{L} \right)
\right]
\,,
\label{eq:fourie}
\end{eqnarray}
where
\begin{equation}
\bar{a}_k 
= 7.56 \times 10^{-5} {\rm eV}^2 
\left(   \frac{\bar{\rho}_k}{{\rm g / cm}^3} \right)
\left(   \frac{E}{{\rm GeV}} \right)
\end{equation}
from eqs.~(\ref{eq:fourie-rho}) and (\ref{eq:mattereffect}).
We further divide $H_1$ to the location independent part
and the location dependent part \cite{koike98}:
\begin{subequations}
\begin{eqnarray}
H_1(x) &=& \overline{H}_1 + \delta H_1(x) \,,
\label{eq:dev-h1}
\\
\overline{H}_1 &=& \frac{1}{2E}U
\left(
\begin{array}{ccc}
0 &  & \\
 & m^2_2 - m^2_1 & \\
 &  & 0
\end{array}
\right)U^{\dagger}
 + \frac{1}{2E} \left(
 \begin{array}{ccc}
\bar{a}_0 &  & \\
 & 0 & \\
 &  & 0
\end{array}
 \right) \,,
 \\
\delta H_1(x) & =&  \sum^{\infty}_{k = 1} 
\frac{1}{E}
\left[ {\rm Re} 
\left(\bar{a}_k \right) 
\cos \left( \frac{2 \pi kx}{L} \right) 
+ {\rm Im} \left( \bar{a}_k  \right)
\sin \left( \frac{2 \pi kx}{L} \right)
\right] 
\left(
\begin{array}{ccc}
1 &   &  \\
  & 0 &  \\
  &   & 0
\end{array}
 \right) 
\,.
\end{eqnarray}
\end{subequations}
\par
The $\nu_{\alpha} \to \nu_{\beta}$ transition matrix elements can now be expressed as
\begin{subequations}
\label{eq:s-total}
\begin{eqnarray}
S(L)_{\beta \alpha}&=&
\left \langle \nu_{\beta} \right |
Te^{-i\int^{L}_{0} (H_0 + H_1(x)) dx}
\left| \nu_{\alpha} \right \rangle
\approx
\left(S_0(L) + S_1(L) + S_2(L) \right ) _{\beta \alpha} 
\label{eq:smatrix}
\,,
\\
S_0(L)_{\beta \alpha} &=& 
\left \langle \nu_{\beta} \right |
e^{-iH_0 L}
\left| \nu_{\alpha} \right \rangle
\label{eq:def-s0}
\,,\\
S_1(L)_{\beta \alpha} &=& -i
\left \langle \nu_{\beta} \right |
\left( \int^{L}_{0} dx e^{iH_0 (x-L)} H_1(x) e^{-iH_0 x} \right)
\left| \nu_{\alpha} \right \rangle
\,,
\label{eq:def-s1}
\\
S_2(L)_{\beta \alpha} &=& -
\left \langle \nu_{\beta} \right |
\left( \int^{L}_{0} dx
\int^{x}_{0} dy e^{iH_0 (x-L)} \overline{H}_1
e^{iH_0 (y-x)} \overline{H}_1 e^{-iH_0 y} \right)
\left| \nu_{\alpha} \right \rangle
\,.
\label{eq:def-s2}
\end{eqnarray}
\end{subequations}
Here we retain only the constant part $\overline{H}_1$
of the perturbation in the second-order term, \eqref{eq:def-s2},
because the higher Fourier modes contribute negligibly
in this order.
The analytic expressions for the above amplitudes
after integration are shown in the appendix.
\par
For the $\nu_{\mu}$ survival mode, $\alpha = \beta = \mu$, we find
\begin{subequations}
\label{eq:s-mu-all}
\begin{eqnarray}
S_0(L)_{\mu \mu} &=& 1 + |U_{\mu 3}|^2 \left( e^{-i \Delta_{13}} - 1\right) \,,
\label{eq:smu-0} \\
S_1(L)_{\mu \mu} &=& -i |U_{\mu 2}|^2 \Delta_{12}
-i |U_{e3}|^2|U_{\mu 3}|^2 \frac{\bar{a}_0 L}{2E} \left( 1 + e^{-i \Delta_{13}}\right)
\nonumber \\
&&- 2|U_{e3}|^2|U_{\mu 3}|^2
\frac{1}{m^2_3 - m^2_1}
\left(
\bar{a}_0 + \sum^{\infty}_{k =1}\frac{ 2 {\rm Re}(\bar{a}_k)}{1 - 4 k^2 (\pi/\Delta_{13})^2}
\right)
\left(
e^{-i\Delta_{13}} - 1
\right)
\label{eq:smu-1}\,, ~~~~~~~~\\
S_2(L)_{\mu \mu} &=& - \frac{\Delta_{12}^2}{2}|U_{\mu 2}|^2
+ {\it O}\left( a^2, a\Delta_{12} \right)
\label{eq:smu-2} \,,
\end{eqnarray}
\end{subequations}
where the phase $\Delta_{ij}$ is
\begin{equation}
\Delta_{ij} = \frac{m^2_j - m^2_i}{2E} L \simeq 2.534
\frac{ (m^2_j - m^2_i) [{\rm eV^2}]}{E {\rm [GeV }]} L [\km ] \,.
\label{eq:Deltaij}
\end{equation}
In \eqref{eq:smu-2}, we ignore the matter effect terms
proportional to $(\bar{a}_0)^2$
because they are proportional to $|U_{e3}|^2$,
which is constrained to be smaller than 0.044 for $|m^2_3 - m^2_1| = 2.35 \times 10^{-3} {\rm eV^2}$
by the CHOOZ experiment \cite{chooz},
and hence can be safely neglected in the second-order terms.
The Fourier terms, $\bar{a}_k$ in \eqref{eq:smu-1}, come along with
the factor of $1/(1 - 4k^2 (\pi/\Delta_{13})^2)$,
which suppresses
contributions of the large $k$ modes.
\par
The $\nu_{\mu}$ survival probability can now be estimated as
\begin{eqnarray}
\hspace{-1.8cm}
P_{\nu_{\mu} \to \nu_{\mu}} (L) &\approx& |S_0(L)_{\mu \mu}+S_1(L)_{\mu \mu} + S_2(L)_{\mu \mu}|^2 
\nonumber \\
&&\hspace{-0.45cm}=
1 - 4 |U_{\mu 3}|^2 \left( 1 - |U_{\mu 3}|^2 \right)
\left ( 
1 + A^{\mu}
\right)
\sin ^2 \left( \displaystyle\frac{\Delta_{13}}{2} + B^{\mu}  \right)
+ {\it O} \left( a^2,~a \Delta_{12}, \Delta_{12}^2 \right) \,,
\label{eq:p-numu-numu}
\end{eqnarray}
where $A^{\mu}$ and $B^{\mu}$ are the correction terms to the amplitude and the oscillation phase,
respectively;
\begin{subequations}
\label{eq:AB-mu}
\begin{eqnarray}
\label{eq:A-mu}
&&\hspace{-1.2cm}
A^{\mu} = - \displaystyle\frac{2}{m_3^2 - m_1^2}
             \left(
                 \bar{a}_0 + \sum^{\infty}_{k=1} 
                  \frac{2 {\rm Re}(\bar{a}_k)}{1 - 4 k^2 (\pi/\Delta_{13})^2}     
             \right)
             \frac{ |U_{e3}|^2 \left( 1 - 2 |U_{\mu 3}|^2 \right) }{1 - |U_{\mu 3}|^2 }
        - \frac{\Delta_{12}^2}{2}\frac{|U_{\mu2}|^2}{1-|U_{\mu 3}|^2}\,,
\\                  
\label{eq:B-mu}
&&\hspace{-1.2cm}B^{\mu} =  \displaystyle\frac{\bar{a}_0L}{4E}\frac{ |U_{e3}|^2 \left( 1 - 2 |U_{\mu 3}|^2 \right) }{1 - |U_{\mu 3}|^2 }  - \displaystyle\frac{\Delta_{12}}{2}\frac{|U_{\mu2}|^2}{1-|U_{\mu 3}|^2} \,.
\end{eqnarray}
\end{subequations}
In the limit of $A_{\mu} = B_{\mu} = 0$, the expression (\ref{eq:p-numu-numu})
reduces to the two-flavor oscillation probability in the vacuum.
\par
In order to show typical orders of magnitude of these correction terms,
we evaluate them for the allowed range of the parameters
\cite{atomos-result, k2k, minos, sol-result, kamland, chooz}:
\begin{subequations}
\label{notation}
\begin{eqnarray}
|m^2_3 - m^2_1| &=& \left(2.35^{+0.11}_{-0.08} \right) \times 10^{-3} {\rm eV}^2 \label{eq:m^2_12bounds}\,, \\
m^2_2 - m^2_1 &=& \left(7.50^{+0.19}_{-0.20} \right) \times 10^{-5 } {\rm eV}^2 \,, \label{eq:m^2_13bounds} \\
4|U_{\mu3}|^2 \left(1 - |U_{\mu3}|^2\right) &=& \sin^2 2\theta_{\rm \footnotesize ATM} > 0.90 \,, \label{eq:sin_atmbounds} \\
4|U_{e1} U_{e2}|^2 &=& \sin^2 2\theta_{\rm \footnotesize SOL} \equiv  0.852^{+0.024}_{-0.026} \,, \label{eq:sin_solbounds}\\
4|U_{e3}|^2\left(1-|U_{e3}|^2\right) &=& \sin^2 2\theta_{\rm \footnotesize RCT} < 0.17 
\mbox{{~  for  ~}} | m_3^2 - m_1^2| = 2.35 \times10^{-3}\mbox{{eV}}^2\,,  \label{eq:sin_rctbounds}
\end{eqnarray}
\end{subequations}
where in eqs. (\ref{eq:sin_atmbounds}) and (\ref{eq:sin_rctbounds}),
the $90\%$ CL bounds are shown.
From eqs.~(\ref{eq:m^2_12bounds}), (\ref{eq:m^2_13bounds}), and (\ref{eq:sin_solbounds}),
we can set
\begin{eqnarray}
\Delta_{12} &=& 0.032 |\Delta_{13}|  \label{eq:Delta13vsDelta12} \,,
\end{eqnarray}
with 5 $\%$ accuracy,
and the allowed range of $|U_{\mu3}|^2$ and $|U_{e3}|^2$ can be parametrized as
\begin{subequations}
\label{eqs:angle-parametrization}
\begin{eqnarray}
\left|U_{\mu3}\right|^2 &\equiv& \sin^2 \theta_{\rm \footnotesize ATM} =0.5 + 0.1 x\,, 
\label{eq:atm-parametrization} \\
\left|U_{e3}\right|^2 &\equiv& \sin^2 \theta_{\rm \footnotesize RCT}  = 0.04y
\label{eq:rct-parametrization} \,,
\end{eqnarray}
\end{subequations}
with two real parameters $x$ and $y$ which can take values
\begin{subequations}
\label{eqs:parameterrization-bound}
\begin{eqnarray}
|x| &=& \left| \displaystyle\frac{|U_{\mu3}|^2 - 0.5}{0.1} \right| < 1.6
 \,, \label{eq:xbounds} \\
y &=& \frac{ \left| U_{e3}\right|^2} {0.04}  < 1.1 \,,
\label{eq:ybounds}
\end{eqnarray}
\end{subequations}
in the 90$\%$ CL allowed regions of eqs.~(\ref{eq:sin_atmbounds}) and (\ref{eq:sin_rctbounds}).
\par
We now obtain the following expressions for $A^{\mu}$ and $B^{\mu}$
around the oscillation maximum, $|\Delta_{13}| \sim \pi$:
\begin{subequations}
\label{eq:AB-mu+}
\begin{eqnarray}
A^{\mu} & \sim &
- 0.0017 \displaystyle\frac{\pi}{\Delta_{13}}\displaystyle\frac{L}{295{\rm km}}\left( \frac{\bar{\rho}_0}{3\density} \right)
\left[ 1 - \left( 0.67 \frac{{\rm Re} (\bar{\rho}_1)}{\bar{\rho_0}} \right) \right]
xy
- 0.0034 \left(\displaystyle\frac{\Delta_{13}}{\pi}\right)^2
\,, ~~~~~~~   \label{eq:A-mu+}   \\
B^{\mu} &\sim& -
0.033 \left[ 1 - 0.23
\sqrt{y}
\cos \dmns 
\right]
\displaystyle\frac{|\Delta_{13}|}{\pi} \,.
\label{eq:B-mu+}
\end{eqnarray}
\end{subequations}
From \eqref{eq:A-mu+}, we find that the 
main uncertainty in
the oscillation amplitude
comes from the sign of $m^2_3 - m^2_1$
or $\Delta_{13}$ 
in the term which is proportional to the matter density
and the baseline length.
Its magnitude around
the first oscillation maximum,
$|\Delta_{13}| \sim \pi$,
is constrained to be less than $0.26\%$ for $\bar{\rho}_0 = 2.62 \density$ and
Re$\bar{\rho}_1/\bar{\rho_0} = 0.015$ for the T2K experiment; see Table 2.
This is about 4 times smaller than
the proposed
sensitivity of $1\%$ for $\sin^2 2\theta_{\rm \footnotesize ATM}$ in
the T2K experiment \cite{T2K}.
\par
As for the corrections to the oscillation phase,
the phase-shift $B^{\mu}$ can be as large as 0.04
for $\delta = \pi$ and $ y = 1.1$
($\sin^2 2\theta_{\rm RCT} = 0.17$) at $|\Delta_{13}| = \pi$.
Although the T2K experiment is expected to measure
the location of the oscillation maximum
$\left| \Delta_{13}/2 + B^{\mu} \right| = \pi/2$ very precisely
with the proposed error of
$0.4 \%$ for $|m^2_3 - m^2_1|$,
the magnitude of $|m^2_3 - m^2_1|$
should depend on the mass hierarchy which can differ from
by as much as $3\%$
\cite{t2kk-full}.
\par
For the $\nu_{\mu} \to \nu_e$ oscillation,
$S_0$ and $S_1$ for the $\nu_{\mu} \to \nu_e$ mode can be expressed as
\begin{eqnarray}
\left[S_0(L)+S_1(L)\right]_{e \mu} 
&=&
\left[
1+
\frac{1}{m^2_3 - m^2_1}
\left\{
\left(
\bar{a}_0 + \sum^{\infty}_{k =1}{\rm Re}(\bar{a}_k)\frac{2 \Delta_{13}^2}{\Delta_{13}^2 - 4 k^2 \pi^2}
\right)
\left(1- 2 |U_{e3}|^2 \right)
\right.
\right. \nonumber \\
&&
\left.
\left.
-i
\sum^{\infty}_{k =1}{\rm Im}(\bar{a}_k)\frac{4 \pi k \Delta_{13}}{\Delta_{13}^2 - 4 k^2 \pi^2}
\right\}
\right]
U_{e3}U_{\mu 3}^*
\left(e^{-i \Delta_{13}} - 1 \right)
\nonumber \\
&&
+ i U_{e3}U_{\mu 3}^* \frac{\bar{a}_0L}{2E}
\left[
1 - |U_{e3}|^2 \left( 1 + e^{-i\Delta_{13}} \right)
\right]
-iU_{e2}U_{\mu 2}^* \Delta_{12} \,.
\label{eq:S0-mue}
\end{eqnarray}
The remarkable point is that the contribution of the Im$(\bar{a}_k)$
terms to the transition probabilities
$|S_0 + S_1|^2$ is highly suppressed,
because 
they cannot interfere with the leading $S_0$ term and
hence they contribute to the probability only as terms of
order $\bar{a}_0$ or $\Delta_{12}$.
The oscillation probabilities
are hence insensitive to the asymmetry
in the matter density distribution
when the sub-leading phase, $\Delta_{12}$ is small.
We confirm this observation quantitatively
in the numerical analysis
and
neglect all terms  of Im $(\bar{a}_k)$ in the following.
\par
The transition probability can then be expressed as
\begin{eqnarray}
P_{\nu_{\mu} \to \nu _e}(L)
&\approx&
\left|S_0(L)_{e \mu} +S_1(L)_{e \mu} \right|^2 + 2{\rm Re}
\left (S_0(L)_{e \mu} S_2^*(L)_{e \mu}  \right)
\nonumber \\
&=&
4 |U_{\mu 3}|^2 |U_{e3}|^2
\left\{
\left( 1 + A^{e} \right)
\sin^2 \left( \displaystyle\frac{\Delta_{13}}{2}\right)
+ B^{e} \sin \left( \Delta_{13} \right)
\right\}
+ C^{e} 
\,,
\label{eq:p-numu-nue}
\end{eqnarray}
where $A^e$ and $B^e$, and $C^e$ are the correction terms;
\begin{subequations}
\label{eq:abc-e}
\begin{eqnarray}
A^e &=& \displaystyle\frac{2\bar{a}_0}{m^2_3 - m^2_1}
\left( 1 - 2|U_{e3}|^2 \right)
\left(
1 + \sum^{\infty}_{k=1}\frac{ 2{\rm Re}\left(\bar{a}_k/\bar{a}_0\right) }{1 - 4 k^2 \left(\pi/\Delta_{13}\right)^2}     
\right)-\displaystyle\frac{1}{2}\left( \displaystyle\frac{\bar{a}_0L}{2E} \right)^2
\left(1 -  \displaystyle\frac{6}{\Delta_{13}^2}  \right)
\label{eq:a-e} \\
&&\hspace{-1.2cm}
+\Delta_{12}\displaystyle\frac{{\rm Im}\left[U_{e2}U_{\mu2}^* U_{e3}^* U_{\mu3} \right]}
{|U_{\mu3}|^2|U_{e3}|^2}
\left(1 + \displaystyle\frac{\bar{a}_0}{\left( m^2_3 - m^2_1\right)}
\right)
+\displaystyle\frac{\Delta_{12}^2}{2}
 \displaystyle\frac{{\rm Re}\left[U_{e2}U_{\mu2}^* U_{e3}^* U_{\mu3} \right]}
{|U_{\mu3}|^2|U_{e3}|^2}
\left( 1 + \displaystyle\frac{\bar{a}_0}{m_2^2 - m_1^2} \right)
\,, \nonumber
\\
B^e &=&
-\displaystyle\frac{\bar{a}_0L}{4E}\left( 1- 2|U_{e3}|^2 \right)
\left(
1 + 2
\displaystyle\frac{\bar{a}_0}{m_3^2 - m_1^2}
\right)
\label{eq:b-e} \\
&&\hspace{-1.2cm}+\displaystyle\frac{\Delta_{12}}{2}
\displaystyle\frac{{\rm Re}\left[U_{e2}U_{\mu2}^* U_{e3}^* U_{\mu3} \right]}
{|U_{\mu3}|^2|U_{e3}|^2}
\left(
1 + \displaystyle\frac{\bar{a}_0}{m^2_3 - m^2_1}
\right)
+\displaystyle\frac{\Delta_{12}^2}{4}
\displaystyle\frac{{\rm Im}\left[U_{e2}U_{\mu2}^* U_{e3}^* U_{\mu3} \right]}
{|U_{\mu3}|^2|U_{e3}|^2}
\left(1 + \displaystyle\frac{\bar{a}_0}{m^2_2 - m^2_1}\right)\,,
\nonumber \\
C^e &=&\Delta_{12}^2|U_{e2}|^2|U_{\mu 2}|^2
        - 2 \Delta_{12}\displaystyle\frac{\bar{a}_0L}{2E}
         {\rm Re}\left[U_{e2}U_{\mu2}^* U_{e3}^* U_{\mu3}\right]
+\left(
\displaystyle\frac{\bar{a}_0 L}{2E}
\right)^2
|U_{e3}|^2
|U_{\mu 3}|^2
\,.
\label{eq:C-term}         
\end{eqnarray}
\end{subequations}
When we substitute the present experimental constraints
of eqs.~(\ref{notation}) - (\ref{eqs:angle-parametrization}),
they can be expressed as
\begin{subequations}
\label{eq:abc-e+}
\begin{eqnarray}
A^e &\sim& 0.11
\displaystyle\frac{\pi}{\Delta_{13}}
\displaystyle\frac{L}{295{\rm km}}
\displaystyle\frac{\bar{\rho}_0}{3 \density}
\left( 1 -0.08 y \right)
\left(
1 + \sum^{\infty}_{k=1}
\displaystyle\frac{2 {\rm Re}\left(\bar{\rho}_1/\bar{\rho}_0\right)  }{1 - 4k^2 \left(\pi/\Delta_{13}\right)^2}
\right)
\nonumber \\
&&
-0.014
\left(\displaystyle\frac{L}{295{\rm km}}
\displaystyle\frac{\bar{\rho}_0}{3 \density}\right)^2
\left(
1
-
0.61
\left(\displaystyle\frac{\pi}{\Delta_{13}}
\right)^2
\right)
\nonumber \\
&&
-0.29
\displaystyle\frac{\sin \dmns}{\sqrt{y}}
\displaystyle\frac{|\Delta_{13}|}{\pi}
\left(1
+0.054 \displaystyle\frac{\pi}{\Delta_{13}}
\displaystyle\frac{L}{295{\rm km}}
\displaystyle\frac{\bar{\rho}_0}{3 \density}
\right)
\nonumber \\
&&
+0.014 \left(
\displaystyle\frac{\cos \dmns}{\sqrt{y}}
- 0.11
\right)
\left(\displaystyle\frac{\Delta_{13}}{\pi}\right)^2
\left(
1 +
1.7
\displaystyle\frac{\pi}{|\Delta_{13}|}
\displaystyle\frac{L}{295{\rm km}}
\displaystyle\frac{\bar{\rho}_0}{3 \density}
\right)
\,,
\label{eq:A-e+} \\
B^e &\sim&
- 0.084
\displaystyle\frac{L}{295{\rm km}} 
\displaystyle\frac{\bar{\rho}_0}{3 \density}
\left(
1 - 0.08 y + 0.11
\displaystyle\frac{\pi}{\Delta_{13}}
\displaystyle\frac{L}{295{\rm km}}
\displaystyle\frac{\bar{\rho}_0}{3 \density}
\right)
\nonumber \\
&&
+0.14
\left(
\displaystyle\frac{ \cos \dmns}{\sqrt{y}}
- 0.12
\right) 
\displaystyle\frac{|\Delta_{13}|}{\pi}
\left(
1 + 0.054
\displaystyle\frac{\pi}{\Delta_{13}}
\displaystyle\frac{L}{295{\rm km}}
\displaystyle\frac{\bar{\rho}_0}{3 \density}
 \right)
\nonumber \\
&&
+0.0072
\displaystyle\frac{\sin \dmns}{\sqrt{y}}
\left(\displaystyle\frac{\Delta_{13}}{\pi}\right)^2
\left(
1 +
1.7
\displaystyle\frac{\pi}{|\Delta_{13}|}
\displaystyle\frac{L}{295{\rm km}}
\displaystyle\frac{\bar{\rho}_0}{3 \density}
\right)
\label{eq:B-e+}\,, \\
C^e &\sim&
0.0011
\left(\displaystyle\frac{\Delta_{13}}{\pi} \right)^2
-
0.0013\sqrt{y}\cos \dmns
\displaystyle\frac{|\Delta_{13}|}{\pi}
\displaystyle\frac{L}{295 {\rm km}}
+
0.00036y
\left(
\displaystyle\frac{L}{295{\rm km}}
\right)^2
\,,
\label{eq:C+-term}         
\end{eqnarray}
\end{subequations}
where we set $x = 0$ $(|U_{\mu3}|^2 = 0.5) $ in the correction factors above for
the sake of brevity.
Finite values of $x$ in the allowed values of
$|x| < 1.6$ do not affect the following discussions significantly \cite{t2kk-atm}.
When the magnitudes of $A^e$ and $B^e$ are both much
smaller than unity, 
these two terms can be factorized as \eqref{eq:p-numu-numu}
for the $\nu_{\mu} \to \nu_{\mu}$ probability.
Therefore
$A^{e}$ 
affects the amplitude of the oscillation and
$B^e$ gives 
the oscillation
phase shift from $\Delta_{13}/2$, except for the small term $C^e$, which does not vanish in the limit of
$|U_{e3}|^2 \to 0$.
It is clear from the above parametrization
that the magnitude of $A^e$ and $B^e$ can be as large
as 0.5 or bigger for $L \simgt  1000$ km,
and we find the expression \eqref{eq:p-numu-nue}
more accurate than the factorized form.
\par
The first term in $A^e$ in \eqref{eq:A-e+}
gives the
matter effect,
whose sign depends on
the mass hierarchy pattern.
When the hierarchy is normal (inverted),
the magnitude of the $\nu_{\mu} \to \nu_e~$ transition probability
is enhanced (suppressed) by
$10 \%$ at Kamioka, and
by 30 $\%$ at $L \sim 1000$ km in Korea
around the oscillation maximum,
$|\Delta_{13}| \sim \pi$.
If we measure the maximum of the oscillation
probability $\sim 4|U_{\mu 3}|^2|U_{e3}|^2 (1+A^e_{\rm peak})$ at two locations, we 
should expect
\begin{eqnarray}
\label{eqs:amplitude-differences}
A^e_{\rm peak}(L\sim1000{\rm km}) - A^e_{\rm peak}(L = 295 {\rm km}) \sim
\left\{
\begin{array}{l}
+0.2 {\rm~for~the~normal~hierarchy}\,, \\ 
-0.2 {\rm ~for~the~inverted~hierarchy} \,,
\end{array}
\right.
\end{eqnarray}
within the allowed range of
the model parameters.
Therefore, simply by comparing the magnitude of the
oscillation maximum between SK and a detector at $L \sim 1000$ km,
we can determine the neutrino mass hierarchy \cite{t2kk-full}.
Once the sign of $\Delta_{13}$ is fixed, $\sin \delta$
can be determined via the third term of \eqref{eq:A-e+}.
\par
Fourier modes of the matter distribution affect
the magnitude of the first term in $A^e$,
and hence affects the $\sin \delta$
measurement.
Since the factor $1/\left( 1 - 4k^2(\pi/\Delta_{13})^2 \right)$
is negative around the first oscillation maximum,
negative ${\rm Re}(\bar{\rho}_k)$ enhances the average matter effect,
and positive ${\rm Re}(\bar{\rho}_k)$ suppresses it.
Figs.~\ref{fig:fourie-295}, \ref{fig:f-real}, and Table~\ref{fig:prem}
show that ${\rm Re} (\bar{\rho}_1)$
is positive 
for the Tokai-to-Kamioka baseline,
while it is negative for the Tokai-to-Korea baselines; see \tableref{tab:compare-prem}.
Hence the first Fourier mode reduces the average matter effect
very slightly by $\sim 1\%$
for the T2K experiment, while
it enhances the effect by about $1 \sim 5\%$ for the Tokai-to-Korea baselines
at around $|\Delta_{13}| \sim \pi$.
Contributions from the higher Fourier modes are highly suppressed.
\par
The term 
$B^e \sin \Delta_{13}$ in \eqref{eq:p-numu-nue} is also sensitive to the sign
of $\Delta_{13}$, and hence the mass hierarchy.
Although its sign depends on the
magnitude of $\cos \delta/\sqrt{y} = \cos \delta/|U_{e3}|$
in the second term of \eqref{eq:B-e+},
the difference in its magnitude at two locations satisfy
\begin{equation}
\label{eq:difference-Be}
B^e\left( L\sim 1000 {\rm km} \right) - B^e\left( L\sim 295 {\rm km} \right)
\sim - 0.2 \,,
\end{equation}
within the allowed range of the model parameters.
If we define the phase of the oscillation maximum by
\begin{equation}
\label{eq:maximum-phase}
\left| \Delta_{13}^{\rm peak}(L) + 2B^e(L) \right| = \pi \,,
\end{equation}
we should expect
\begin{eqnarray}
\label{eqs:phase-differences}
\Delta_{13}^{\rm peak}(L\sim1000{\rm km}) - \Delta_{13}^{\rm peak}(L)(L = 295 {\rm km})\sim
\left\{
\begin{array}{l}
+0.4 {\rm~for~the~normal~hierarchy}, 
\\
-0.4 {\rm ~for~the~inverted~hierarchy} \,.
\end{array}
\right.
\end{eqnarray}
This implies that the mass hierarchy can be determined also by
measuring the location of the $\nu_{\mu} \to \nu_e$ oscillation
maximum with the accuracy better than around $10\%$.
\par
It should also be noted that the
terms proportional to $\sin \delta$ and $\cos \delta$
in eqs. (\ref{eq:A-e+}) and (\ref{eq:B-e+})
are proportional to $1/\sqrt{y} \propto 1/|U_{e3}|$.
Because the $\nu_\mu \to \nu_e$ oscillation probability 
is proportional to $y \propto |U_{e3}|^2$ in \eqref{eq:p-numu-nue},
the statistical error of its measurement is proportional to $\sqrt{y}$.
The $1/\sqrt{y}$ factor in front of the
$\sin \delta$ and $\cos \delta$
terms cancel this $\sqrt{y}$ factor,
and hence the error of the $\delta$ measurements does not depend
on the magnitude of $|U_{e3}|^2$, as long as the
neutrino mass hierarchy is determined \cite{t2kr-l, t2kk-full}.
\par
From the above observations, we understand quantitatively
the reason why the mass hierarchy and both $\sin \delta$ and $\cos \delta$
can be measured by observing \mutoe oscillation around the first oscillation maximum
at two vastly different baseline lengths.
The T2KK proposal of Refs.~\cite{t2kr-l, t2kk-full} showed that this
goal can be achieved by using the existing detector SK at $L = 295$ km
and only one new detector at a specific location in the Korean east coast
where the beam can be observed at an off-axis angle
below $1^{\circ}$, so that the neutrino flux at the first oscillation maximum
$(\sim$ a few GeV) is significant.
\par
In this report, we investigate further the impacts of
using $\bar{\nu}_{\mu}$ beam in addition to ${\nu}_{\mu}$ beam.
The oscillation probabilities $P(\bar{\nu}_\mu \to \bar{\nu}_\mu$)
and $P(\bar{\nu}_\mu \to \bar{\nu}_e$) are obtained
from the expressions for $P(\nu_\mu \to \nu_\mu$) and
$P(\nu_\mu \to \nu_e$), respectively,
by reversing the sign of the matter effect term ($\rho \to -\rho$)
and that of $\delta$.
Because these sign changes occur independently
of the $L$-dependence of the oscillation probabilities,
we may expect improvements in the physics potential
of the T2KK experiment for a given number of POT.
\subsection{Hybrid method}
In this subsection,
we introduce a very simple approximation formula for the oscillation
probabilities that can be used for fast simulation purposes.
The analytic formula of the previous subsection are useful in understanding qualitatively
the dependences of the oscillation probabilities
on the model parameters and the baseline lengths,
but they are not suited for quantitative analysis presented below.
It is because the approximation of keeping only the
first and second orders of
the matter effect terms
$\bar{\rho}_i$ and the
non-leading oscillation phase  $\Delta_{12}$ is not 
excellent at far distances ($L \sim 1000$ km), as
can be inferred from the magnitude of
the correction terms $A^e$ and $B^e$ in eq.~(\ref{eq:abc-e}).
\par
On the other hand, the average and Fourier coefficients of the matter density profile
shown
in \tableref{tab:compare-prem}
for Tokai-to-Korea baselines tell that Re($\bar{\rho}_1$)/$\bar{\rho}_0$ is
rather small, $1.5\%$ for T2K and $1 \sim 5\%$ for Tokai-to-Korea baselines
$( L = 1000 \sim 1200 {\rm km})$.
Higher Fourier modes are expected to contribute even less. 
Therefore, we expect that
the perturbation in terms of ${\rm Re}(\bar{a}_1)$ to
the oscillation probabilities for the uniform matter density,
$\rho(x) = \bar{\rho}_0$,
can give a good approximation.
The $\nu_{\mu} \to \nu_{\mu}$ survival and
$\nu_{\mu} \to \nu_e$ transition probabilities in such approximation
are expressed as
\begin{subequations}
\label{eq:hybrid}
\begin{eqnarray}
&\hspace{-0.7cm}
P^{\rm hybrid}_{\nu_\mu \to \nu_\mu}
\left(\bar{a}_0,~{\rm Re}(\bar{a}_1) \right) = P^{\rm exact}_{\nu_\mu \to \nu_\mu}(\bar{a}_0)
+
\displaystyle\frac{16{\rm Re}(\bar{a}_1)}{m^2_3 - m^2_1}
\frac{\left|U_{e3}U_{\mu 3}\right|^2 \left(1-2\left|U_{\mu3}\right|^2\right) }
{1-4\left(\pi/\Delta_{13}\right)^2}
\sin^2 \left( \displaystyle\frac{\Delta_{13}}{2}\right)
\,,
\label{eq:hy-mumu}
\\
&\hspace{-0.7cm}
P^{\rm hybrid}_{\nu_\mu \to \nu_e}
\left(\bar{a}_0,~{\rm Re}(\bar{a}_1) \right) = P^{\rm exact}_{\nu_\mu \to \nu_e}(\bar{a}_0)
+
\displaystyle\frac{16{\rm Re}(\bar{a}_1)}{m^2_3 - m^2_1}
\frac{|U_{e3} U_{\mu 3}|^2 \left(1-2|U_{e3}|^2\right)}{1-4\left(\pi/\Delta_{13} \right)^2}
\sin^2 \left( \displaystyle\frac{\Delta_{13}}{2}\right)
\,,
\label{eq:hy-mue}
\end{eqnarray}
\end{subequations}
respectively.
We call the approximation hybrid, being a combination of the
exact oscillation probabilities for a constant
matter density and the first order analytic correction term
which is proportional to the real part of the first Fourier coefficient 
of the matter density distribution along the baseline.
\par
In \eqref{eq:hy-mumu}, the second term
is suppressed because $ |1-2|U_{\mu 3}|^2 | = 0.2 x < 0.32$,
see eqs.~(\ref{eqs:angle-parametrization}) and (\ref{eqs:parameterrization-bound}).
We find that \eqref{eq:hy-mumu} gives an excellent approximation
to the $\nu_{\mu} \to \nu_{\mu}$ survival probability in the whole range
of the parameter space explored in this report. Equation (\ref{eq:hy-mue})
also gives an excellent approximation to the
$\nu_{\mu} \to \nu_e$ oscillation probability around the
oscillation maxima. However it does not always
reproduce accurately the probability around the oscillation minima at which
the first term gets very small and our first order correction term 
can be off significantly due to the phase-shift.
We find that the difference between the exact formula
and the hybrid method around the oscillation minimum
can be as large as $10 \sim 30\%$.
Because the number of event is also small around the minimum region,
this difference is not significant for the $\Delta \chi^2$ analysis
in most of the parameter space.
We find that 
the minimum
$\Delta \chi^2$ values are obtained rather accurately
when the exact formula are replaced by the approximation of \eqref{eq:hybrid}.
\section{Analysis method}
Before we present the results of our numerical calculation, we would like to explain
our treatment of the signals and backgrounds, and the analysis method.
\par
In order to compare our results with those of the previous works, we adopt the same setting
of Ref.~\cite{t2kk-full}.
There,
the detectors at both Kamioka and Korea are
assumed to have excellent detection and kinematical reconstruction capabilities
for $\nu_{\mu}$ and $\nu_e$
Charged Current Quasi-Elastic (CCQE) events within
the fiducial volumes of the 22.5 kt at Kamioka (SK) and 100 kt at Korea.
The energy bin width, $\delta E$, in the analysis is chosen as $200$ MeV for  
$E_\nu > 400$ MeV, which
accounts partially for kinematical reconstruction
errors\footnote{See ${\it e.g.}$ Ref.~\cite{t2kk-bg} for a more realistic treatment,
where smearing effects due to nucleonic Fermi motion,
resonance production and finite detector resolutions have been studied.
}.
\par
The $\nu_{\mu} \to \nu_{\alpha}$ oscillation signals in the $i$-th energy bin,
$E_\nu^i = 200{\rm MeV} \times (i+1) < E_{\nu} < E_{\nu}^i + \delta E_{\nu}$,
at each site are calculated as
\begin{equation}
\label{eq:N}
 N_{\alpha,~{\rm D}}^{i} (\nu_\mu)=
 M_{\rm D} N_A
 \int_{E_\nu^i}^{E_\nu^{i}+\delta E_\nu}
 \Phi_{\nu_\mu}(E, L)~
 P_{\nu_\mu\to\nu_\alpha}(E)~ 
 \sigma_\alpha^{CCQE}(E)~
 dE\,,
\end{equation}
where the suffix D = SK or Kr denotes the detector, and
$P_{\nu_{\mu}\rightarrow \nu_{\alpha}}$
is the neutrino oscillation probability,
which is calculated by the exact formula, \eqref{eq:multi-time-evo}.
$M_{\rm SK} = 22.5$ kt and $M_{\rm Kr} = 100$ kt
are the detector mass within the fiducial volume,
$N_{A} = 6.017\times10^{23}$ is the Avogadro number,
$\Phi_{\nu_\mu}$ is the $\nu_{\mu}$ flux from J-PARC \cite{ichikawa}
which is proportional to $1/L^2$,
and
$\sigma_\mu^{CCQE}$ and
$\sigma_e^{CCQE}$
are, respectively, $\nu_{\mu}$ and $\nu_e$ CCQE cross sections per nucleon in water \cite{cross-section}.
\par
As for the background, we only consider the contribution from the secondary neutrino 
fluxes of the  $\nu_\mu$ primary beam, such as $\nu_{e}, \bar{\nu}_{\mu}, \bar{\nu}_e$
which are calculated as in eq.~(\ref{eq:N})
where $\Phi_{\nu_{\mu}}(E)~$ 
and $P_{\nu_{\mu} \to \nu_{\alpha}}$
are replaced by $\Phi_{\nu_{\beta}}(E)$
and $P_{\nu_{\beta} \to \nu_{\alpha}}$,
respectively, for
$\nu_{\beta} = \nu_e\,,\bar{\nu}_e\,,\bar{\nu}_{\mu}$,
and $\nu_{\alpha} = \nu_e$ or $\bar{\nu}_e$ for
$\alpha = e$, and $\nu_{\mu}$ or $\bar{\nu}_{\mu}$
for $\alpha = \mu$.
The data of all the primary and the secondary fluxes
as well as the CCQE cross sections
used in our analysis
can be obtained from the website \cite{ichikawa}.
After summing up these background events,
the numbers of $e$-like and $\mu$-like events in the $i$-th bin are
calculated as
\begin{equation}
\label{signal}
 N_{\alpha,{\rm D}}^{i} = N_{\alpha,{\rm D}}^{i}(\nu_\mu)
 + N_{\alpha,{\rm D}}^{i,{\rm BG}}
\,,
{\mbox{\hspace*{5ex}}}
(\alpha = e, \mu;~{\rm D}={\rm SK,~Kr})\,.
\end{equation}
When the primary $\nu_{\mu}$ beam is replaced by the primary $\bar{\nu}_{\mu}$ beam,
the signals are obtained from $\bar{\nu}_{\mu} \to \bar{\nu}_\alpha \left(\alpha = e, \mu\right)$
while the backgrounds are from $\nu_{\beta} \to \nu_{\alpha}$ with $\nu_{\beta} = \bar{\nu}_e, \nu_e, \nu_{\mu}$.
\par
As a quantitative indicator of the capability of the proposed
experiments to measure the neutrino model parameters,
we introduce a
$\Delta \chi^2$ function as follows:
\begin{equation}
\label{chi^2 define}
\Delta\chi^2 \equiv  \chi^2_{\rm stat} + \chi^2_{\rm sys} + \chi^2_{\rm para}\,.
\end{equation}
The first term, $\chi^2_{\rm stat}$,
accounts for
the parameter dependence of the fit to the CCQE events,
\begin{eqnarray}
\label{eq:chi^2event}
 \chi^2_{\rm stat}
 = \sum_{{\rm D} ={\rm SK, Kr}}
  \sum_{i} \left\{
\left(
\displaystyle\frac
{(N_{e,{\rm D}}^{i})^{\rm fit} - N_{e, {\rm D}}^{i}}
{\sqrt{N^i_{e, {\rm D}}}}
\right)^2
+
\left(
\displaystyle\frac
{(N_{\mu, {\rm D}}^{i})^{\rm fit} - N_{\mu, {\rm D}}^{i}}
{\sqrt{N^i_{\mu, {\rm D}}}}
\right)^2
\right\}\,,
\end{eqnarray}
where the summation is over all bins
from 0.4 GeV to 5.0 GeV for $N_{\mu}$ at both sites,
0.4 GeV to 1.2 GeV for $N_{e, {\rm SK}}$,
and 0.4 GeV to 2.8 GeV for $N_{e, {\rm Kr}}$.
Here $N^i_{e, {\rm D}}$ and $N^i_{\mu, {\rm D}}$ are the calculated number of
$e$-like and $\mu$-like events
in the $i$-th bin in each detector (${\rm D} = {\rm SK, Kr}$),
and their square-roots give the statistical error.
The neutrino oscillation probabilities $P_{\nu_\mu \to \nu_\alpha}(E)$
in \eqref{eq:N} are calculated for
the fallowing input parameters:
\begin{subequations}
\label{eqs:simulation-inputprameters}
\begin{eqnarray}
|m^2_3 - m^2_1| &=& 2.35 \times 10^{-3} {\rm eV}^2 \,, \\
m^2_2 - m^2_1 &=& 7.5 \times 10^{-5 } {\rm eV}^2 \,, \\
|U_{\mu3}|^2 &=& \sin^2 \theta_{\rm \footnotesize ATM} = 0.5 \,, \\
4|U_{e1} U_{e2}|^2 &=& \sin^2 2\theta_{\rm \footnotesize SOL} = 0.85 \,,
\end{eqnarray}
\end{subequations}
for both hierarchies ($m^2_3 - m^2_1 >0$ or $m^2_3 - m^2_1 <0$),
and for various values of $\sin^2 2\theta_{\rm RCT}$ and $\delta$.
For the matter density profile along the baselines, we use
the mean matter density of each regions as given
in Figs.~\ref{fig:t2k-profile} and \ref{fig:t2korea}.
\par
The numbers,
$\left(N^{i}_{e,{\rm D}}\right)^{\rm {\rm fit}}$ and $\left(N^{i}_{\mu,D}\right)^{\rm fit}$
are then calculated by allowing the model parameters
to vary freely
and by allowing for systematic errors.
We consider the following systematic errors in this analysis.
We assign $6\%$ uncertainty to the overall matter density
along each baseline, Tokai-to-Kamioka and
Tokai-to-Korea,
because the ambiguity in the density-velocity conversion scale dominates the uncertainty
of the average matter density:
\begin{eqnarray}
\rho(x)^{\rm T2K,fit}  &=& f_{\rho}^{\rm SK} \rho^{\rm SK}(x) \,,~~~
\rho(x)^{\rm T2Kr,fit}  = f_{\rho}^{\rm Kr} \rho^{\rm Kr}(x)\,,
~~~~f_{\rho}^{\rm SK},~f_{\rho}^{\rm Kr} = 1 \pm 0.06\,. \label{eq:fit-density}
\end{eqnarray}
Although we expect positive correlation between
the scale factors of $f_{\rho}^{\rm SK}$ and $f_{\rho}^{\rm Kr}$,
we treat them independently as a conservative estimate.
We assign $3\%$ uncertainty in the normalizations
of each beam flux:
\begin{eqnarray}
\label{eq:fit-def}
\Phi_{\alpha}^{\rm fit}(E, L) &=& f_{\nu_{\alpha}}\Phi_{\alpha} (E,L)\,,~~~~
f_{\nu_{\alpha}} = 1 \pm 0.03
~~
\left({\rm for}~\nu_{\alpha} = \nu_{\mu},~\bar{\nu}_{\mu},~\nu_e,~\bar{\nu}_e\right)
\,.
\end{eqnarray}
Here also,
we ignore possible correlations among the errors of primary and secondary
beam fluxes.
For the CCQE cross sections of neutrinos and anti-neutrinos,
we assign common $3\%$ error for $\nu_{\mu}$ and $\nu_e$ events;
\begin{equation}
\label{eq:ccqe-normalization-neutrino}
\sigma_{\nu_{\mu}}^{{\rm CCQE, fit}}(E) = f_{l} \sigma_{\nu_{\mu}}^{{\rm CCQE}}(E),
\hspace{0.5cm}
\sigma_{\nu_{e}}^{{\rm CCQE, fit}}(E) = f_{l} \sigma_{\nu_{e}}^{{\rm CCQE}}(E) \,,
~~~~
f_{l} = 1 \pm 0.03\,,
\end{equation}
and another common $3\%$ error for $\bar{\nu}_{\mu}$ and $\bar{\nu}_e$
events;
\begin{equation}
\label{eq:ccqe-normalization-antineutrino}
\sigma_{\bar{\nu}_{\mu}}^{{\rm CCQE, fit}}(E) = f_{\bar{l}} \sigma_{\bar{\nu}_{\mu}}^{{\rm CCQE}}(E),
\hspace{0.5cm}
\sigma_{\bar{\nu}_{e}}^{{\rm CCQE, fit}}(E) = f_{\bar{l}} \sigma_{\bar{\nu}_{e}}^{{\rm CCQE}}(E) \,,
~~~~
f_{\bar{l}} = 1 \pm 0.03\,.
\end{equation}
This is equivalent of assuming the $e$-$\mu$ universality in the cross section
uncertainly, while neglecting correlation between $\nu_l$ and $\bar{\nu}_l$
cross section errors.
For the fiducial volume of SK and a far detector in Korea,
we assign $3\%$ error each:
\begin{eqnarray}
M_{\rm D}^{\rm fit} &=& f_{\rm D} M_{\rm D}\,,
~~~~
f_{\rm D} = 1 \pm 0.03
\hspace{0.3cm} ({\rm for~D} = {\rm SK,~Kr})
\,. 
\end{eqnarray}
Summing up, we introduce 10 normalization errors,
when calculating $(N^i)^{\rm fit}$,
and $\chi^2_{\rm sys}$ is expressed as
\begin{eqnarray}
\hspace{-0.3cm}\chi^2_{\rm sys} &=&
\sum_{\nu_{\alpha} = \nu_{\mu},\bar{\nu}_{\mu}, \nu_e,\bar{\nu}_{e}}
\left(
\displaystyle\frac{f_{\nu_{\alpha}}-1}{0.03}
\right)^2
+
\sum_{\beta = l, \bar{l}}
\left(
\displaystyle\frac{f_{\beta}-1}{0.03}
\right)^2
\nonumber \\
&& \hspace{0.2cm}
+
\sum_{ D = {\rm SK,~Kr}}
\left[
\left(
\displaystyle\frac{f_D-1}{0.03}
\right)^2
+
\left(
\displaystyle\frac{f^{D}_{\rho}-1}{0.06}
\right)^2
\right]\,.
\label{chisq-sys}
\end{eqnarray}
\par
Finally, $\chi^2_{\rm para}$ accounts for external constraints 
on the model parameters:
\begin{eqnarray}
\label{eq:chisq-para}
\chi^2_{\rm para}
&=&
\left(
\displaystyle\frac{(m_2^2 - m_1^2)^{\rm fit} - 7.5 \times 10^{-5} {\rm eV}^2}
{ 0.2 \times 10^{-5}}
\right)^2 +
\left(
\displaystyle\frac
{\sin^22\theta_{\rm \footnotesize SOL}^{\rm fit}- 0.85}
{0.025}
\right)^2 
\nonumber
\\
&&
+
\left(
\displaystyle\frac
{\sin^22\theta_{\rm \footnotesize RCT}^{\rm fit}- \sin^22\theta_{\rm \footnotesize RCT}^{\rm input}}
{0.01}
\right)^2\,.
\end{eqnarray}
The first two terms are essentially the present constraints from the KamLAND experiment
\cite{sol-result},
summarized in \eqref{notation}.
In the last term, we assume that the planned future reactor experiments
\cite{double-chooz, reno, daya-bay}
should measure \lsir with an uncertainty of 0.01, during the T2KK experimental period.
We do not impose the present constraints on
$|m^2_3 - m^2_1|$ and $\sin^2 2\theta_{\rm ATM}$ given in \eqref{notation},
since the sensitivity of the T2KK experiments supersedes them.
\par
In total, $\Delta \chi^2$ depends on six model parameters
and ten normalization factors.
The minimum of $\Delta \chi^2$ is then found in this 16 dimensional parameter space.
\par
\section{The earth matter effects and the mass hierarchy}
In this section,
we discuss the relation between
the earth matter effects and
the capability of determining the neutrino mass hierarchy pattern
in the T2KK experiment.
We also examine the dependence of the capability on the location of
a far detector in Korea.
\subsection{Determining the mass hierarchy}
The analytic expressions presented in the previous
section show that $\nu_{\mu} \to \nu_{\mu}$ and
$\bar{\nu}_{\mu} \to \bar{\nu}_{\mu}$ oscillations
are insensitive to the neutrino mass hierarchy,
while $\nu_{\mu} \to \nu_{e}$ and
$\bar{\nu}_{\mu} \to \bar{\nu}_{e}$
oscillation probabilities depend significantly on
the mass hierarchy pattern through the terms which are proportional
to the baseline length $L$ and the average matter density $\bar{\rho_0}$.
Two independent observables are sensitive to the mass hierarchy.
According to the estimates of eqs.~(\ref{eqs:amplitude-differences})
and (\ref{eqs:phase-differences}), the magnitude of
the first oscillation maximum
at $L \sim 1000$ km should be about $20 \%$ bigger (smaller)
than that of $L = 295$ km
when the mass hierarchy is normal (inverted).
Likewise the location of the phase
where the first peak occurs at $L \sim 1000$ km is
about $25^{\circ}$ bigger (smaller) than
that at $L = 295$ km.
All the above predictions should be reversed for $\bar{\nu}_{\mu} \to \bar{\nu}_e$
oscillation.
It is remarkable that none of the other model parameters
contribute significantly to the differences
eqs.~(\ref{eqs:amplitude-differences}) and (\ref{eqs:phase-differences}),
since they tend to cancel between the two measurements
of the same oscillation probabilities.
\par
Since the matter effect is proportional to the average matter density $\bar{\rho}_0$ 
along each baseline  with corrections from higher Fourier modes,
we expect that capability of constraining the mass hierarchy pattern in the T2KK
experiment depends on the earth matter profile along the baseline.
\par
\begin{table}[t]
\begin{center}
\begin {tabular}{|c|c|c|c|c|} \hline
$\delta$& $0^{\circ}$& $90^{\circ}$ &
$180^{\circ}$ & $270^{\circ}$ \\ \hline
previous results \cite{t2kk-full} & 14.6  & 23.0 & 54.7 &42.9
\\ \hline
changes in input model parameters &	
14.7  & 25.1 & 53.1 &48.7
\\ \hline
overall error of the density ($3 \% \rightarrow 6 \%$) &
14.4  & 24.7  & 51.2  & 47.0 
\\ \hline
average density SK($2.8 \density \rightarrow 2.6 \density$) &
14.6 & 24.9 & 51.8 & 47.0 \\ \hline
average density Kr ($3.00 \density \rightarrow 2.98 \density$) &
14.2 & 24.4 & 51.0 & 47.0\\ \hline
non-uniform matter density profile &
15.7 & 26.8 & 54.2 & 49.1 \\ \hline
\end{tabular}
\end{center}
\caption{
The changes in minimum $\Delta \chi^2$ values at
each modification on the matter effect treatment
from the previous work \cite{t2kk-full}.
Signals are generated
for $\sin^2 2\theta_{\rm RCT} = 0.10$ and the four $\delta$
values by assuming the normal hierarchy
and fit is performed by assuming the inverted hierarchy.
The combination of off-axis angles are $3.0^{\circ}$ at SK
and $1.0^{\circ}$ at Korea, and the Tokai-to-Korea baseline length
is 1100 km. The results are for the neutrino beam with $5 \times 10^{21}$ POT. 
}
\label{tab:pull-matter}
\end{table}
In order to
examine the impacts of the
non-uniform matter profile along the baselines,
let us compare our result with those obtained in Ref.~\cite{t2kk-full},
where
the matter profiles along the baselines are assumed to be uniform
with densities  $2.8 {\rm g/cm^3}$ for T2K
and $3.0 {\rm g/cm^3}$ for Tokai-to-Korea.
The errors of these densities have been assumed to be $3\%$
\cite{t2kr-l, t2kk-full, t2kk-atm}.
We summarize changes in $\Delta \chi^2$ values for 
each modification
step by step
in Table.~\ref{tab:pull-matter}.
Here, the minimum
$\Delta \chi^2$ is 
obtained from the data generated
by assuming the normal hierarchy ($m^2_3 - m^2_1 >0$)
while the fit is performed by assuming the inverted hierarchy ($m^2_3 - m^2_1 <0$).
The input parameters are $\sin^2 2\theta_{\rm RCT} = 0.10$ and
$\delta = n \times 90^{\circ} ( n =~0,~1,~2,~3)$.
The combination of off-axis beam is $3.0^{\circ}$ at SK
and $1.0^{\circ}$ in Korea at $L = 1100$ km.
\par
The first row of \tableref{tab:pull-matter}
gives input $\delta$ values,
and the second row gives the results of Ref.~\cite{t2kk-full}
which are reproduced in our analysis by using the same input parameters.
In the third row, exactly the same computation is repeated when
the input parameter values are updated
according to more recent experimental data, summarized in \eqref{notation}.
It is remarkable that the minimum $\Delta \chi^2$ values
do not change much even though
the errors of $m^2_2 - m^2_1$ and $\sin^2 2\theta_{\rm SOL}$
have been reduced by a factor of three since 2006
when the analysis of Ref.~\cite{t2kk-full} were performed. 
\par
The first modification in the matter effect
is the magnitude of the uncertainty in the matter density,
which changes from $3\%$ in Ref.~\cite{t2kk-full} to
$6\%$ in this report according to discussions in section 2; see \figref{fig:v-rho-trans}.
Although the matter effect is the key to determine the mass hierarchy pattern
in the T2KK experiment,
the larger error of the density does not affect $\Delta \chi^2$ significantly.
This is because the statistical error is dominant in the present setting.
We note in passing that we assign $6\%$ error independently
for T2K ($f^{\rm SK}_\rho$) and Tokai-to-Korea 
($f^{\rm Kr}_\rho$) baselines, even though the dominant systematic error in the conversion
of sound velocity to the matter density is common. This makes our estimates conservative,
since the typical size of the independent error is $\sim 3\%$ \cite{t2kk}. 
\par
The second and the third modifications are the values of the average matter densities.
The reduction of the average matter density for the T2K baseline
due to Fossa Magna improves the mass hierarchy discriminating power
since the differences eqs.~(\ref{eqs:amplitude-differences})
and (\ref{eqs:phase-differences}) increase.
A slightly smaller value of the average matter density
at $L \sim 1100$ km decreases $\Delta \chi^2$: The latter effect is non-negligible
because the differences are proportional to
$1100 \times \bar{\rho}_0 (L = 1100 {\rm km}) - 295  \times \bar{\rho}_0 ({\rm T2K})$.
\par
Final modification is the non-uniformity of the matter 
density distribution.
It is remarkable that the minimum $\Delta \chi^2$ increases
significantly for all four $\delta$ cases,
as consequences of non-uniformity in the matter density distributions.
The effect is particularly significant for T2KK,
because ${\rm Re}(\bar{\rho}_1)$ is {\it positive} for the T2K baseline due
to Fossa Magna;
see \figref{fig:t2k-profile} and \tableref{tab:compare-prem}.
\begin{figure}[t]
\begin{center}
\includegraphics[angle = 0 ,width=6cm]{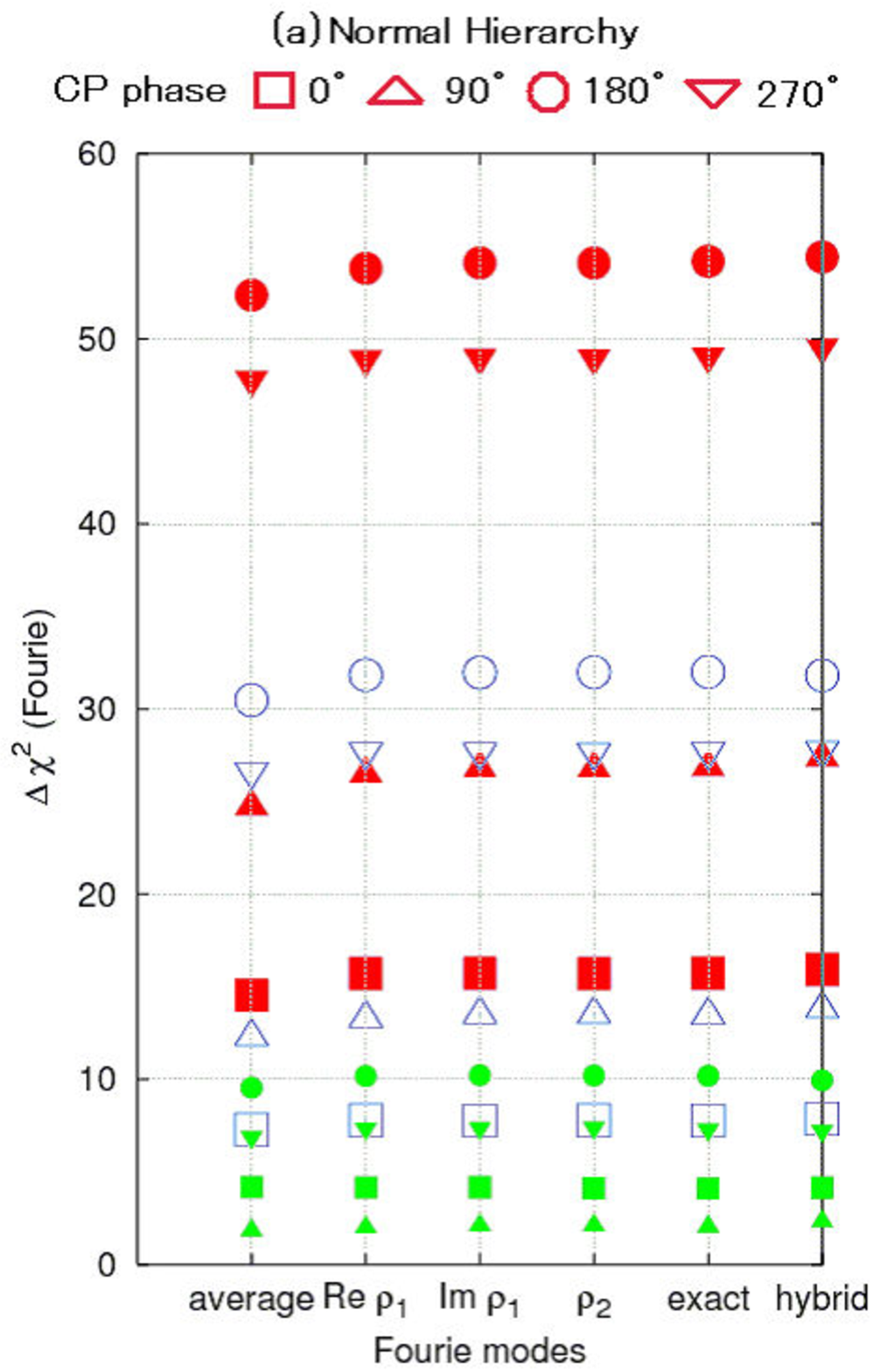}~~~~
\includegraphics[angle = 0 ,width=6cm]{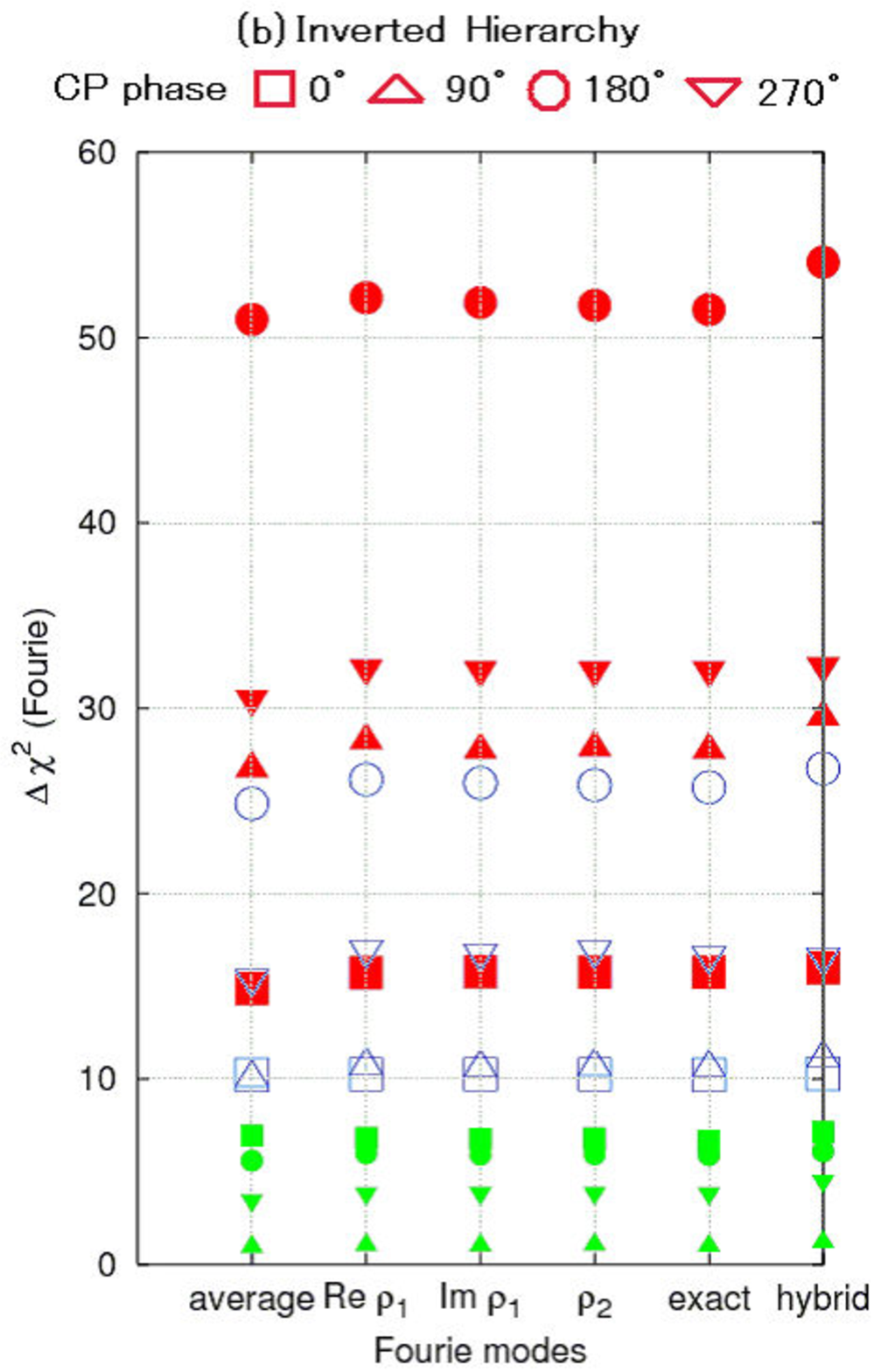}
\end{center}
\caption{Minimum $\Delta\chi^2$ of the T2KK two detector experiment
when the matter distribution along the baselines are
approximated by their average, and up to first and
second Fourier modes.
They are compared with the results of the exact (step-function-like)
matter distribution and those of the hybrid approximation
to the oscillation probabilities.
The input hierarchy is normal (inverted) in the plot a (b) 
where the opposite hierarchy is assumed in the fit.
The input $\sin^2 2\theta_{\rm RCT}$ are 0.10 for solid red,
0.06 for open blue, and 0.02 for solid green symbols,
and 
the input CP phase are $0^{\circ}$ for squares,
$90^{\circ}$ for upper triangles, $180^{\circ}$ for circle,
and $270^{\circ}$ for upside-down triangles.
The results are for $3.0^{\circ}$ OAB at SK, and for a 100 kt
far detector at $L = 1100$ km observing $1.0^{\circ}$ OAB,
with the neutrino flux from $5 \times 10^{21}$ POT at J-PARC.
}
\label{fig:fourie-chi}
\end{figure}
\par 
In \figref{fig:fourie-chi}, we show the minimum values of $\Delta\chi^2$ for the T2KK two detector experiment
as functions of the matter density distributions which are 
approximated by the distribution with the average
density (average), including the real-part of the first Fourier mode
(Re$\bar{\rho}_1$), further including the imaginary part (Im$\bar{\rho}_1$),
and also with the second modes ($\bar{\rho}_2$).
The results are then compared
with $\Delta \chi^2_{\rm min}$ calculated by
using the exact (step-function-like)
distribution.
As a reference,
$\Delta \chi^2_{\rm min}$ calculated
by the hybrid approximation of
the oscillation probability, \eqref{eq:hybrid},
is also shown (hybrid). 
The input hierarchy is normal in the left figure
while it is inverted in the right figure,
where in both cases
the opposite hierarchy is assumed in the fit.
The input $\sin^2 2\theta_{\rm RCT}$ is 0.10 for solid red blobs,
0.06 for open blue symbols, whereas 0.02 for solid green blobs,
whereas
the input CP phase are $0^{\circ}$ for squares,
$90^{\circ}$ for upper triangles, $180^{\circ}$ for circle,
and $270^{\circ}$ for upside-down triangles.
We show the results for a far detector at $L = 1100$ km
baseline observing $1.0^{\circ}$ OAB, which has
$- {\rm Re}(\bar{\rho}_1)/\bar{\rho_0} = 0.036$.
\figref{fig:fourie-chi} shows that the minimum $\Delta \chi^2$ values
are reproduced accurately for all the cases just
by accounting for the Re$(\bar{\rho}_1)$ component.
We conclude that 
higher Fourier modes including Im($\bar{\rho}_1$)
do not affect the minimum $\Delta \chi^2$ significantly in the T2KK experiment.
\par
The approximation of keeping only the linear term in Re$(\bar{\rho}_1)$
in the oscillation probabilities, the hybrid method of section 3.3,
works rather well at most cases.
However, in \figref{fig:fourie-chi}(b),
the hybrid method gives slightly high minimum $\Delta \chi^2$ for
$\delta = 180^{\circ}$ and $90^{\circ}$ at
$\sin^2 2\theta_{\rm RCT} = 0.1$,  
and for $\delta = 180^{\circ}$ at $\sin^2 2\theta_{\rm RCT} = 0.06$.
In these cases,
the hybrid method
overestimates $\Delta \chi^2$ in bins around the oscillation
minimum where the approximation is poor; see discussion
in section 3.3. In more realistic
studies including neutral current $\pi^0$ background, see ${\it e.g.}$ \cite{t2kk-bg},
those bins around the oscillation minima play less
significant role due to the background dominance,
and we expect that the hybrid method gives more reliable results. 
\subsection{Optimal T2KK setting for determining the mass hierarchy}
\begin{figure}[t]
\begin{center}
\includegraphics[angle = 0 ,width=7.5cm]{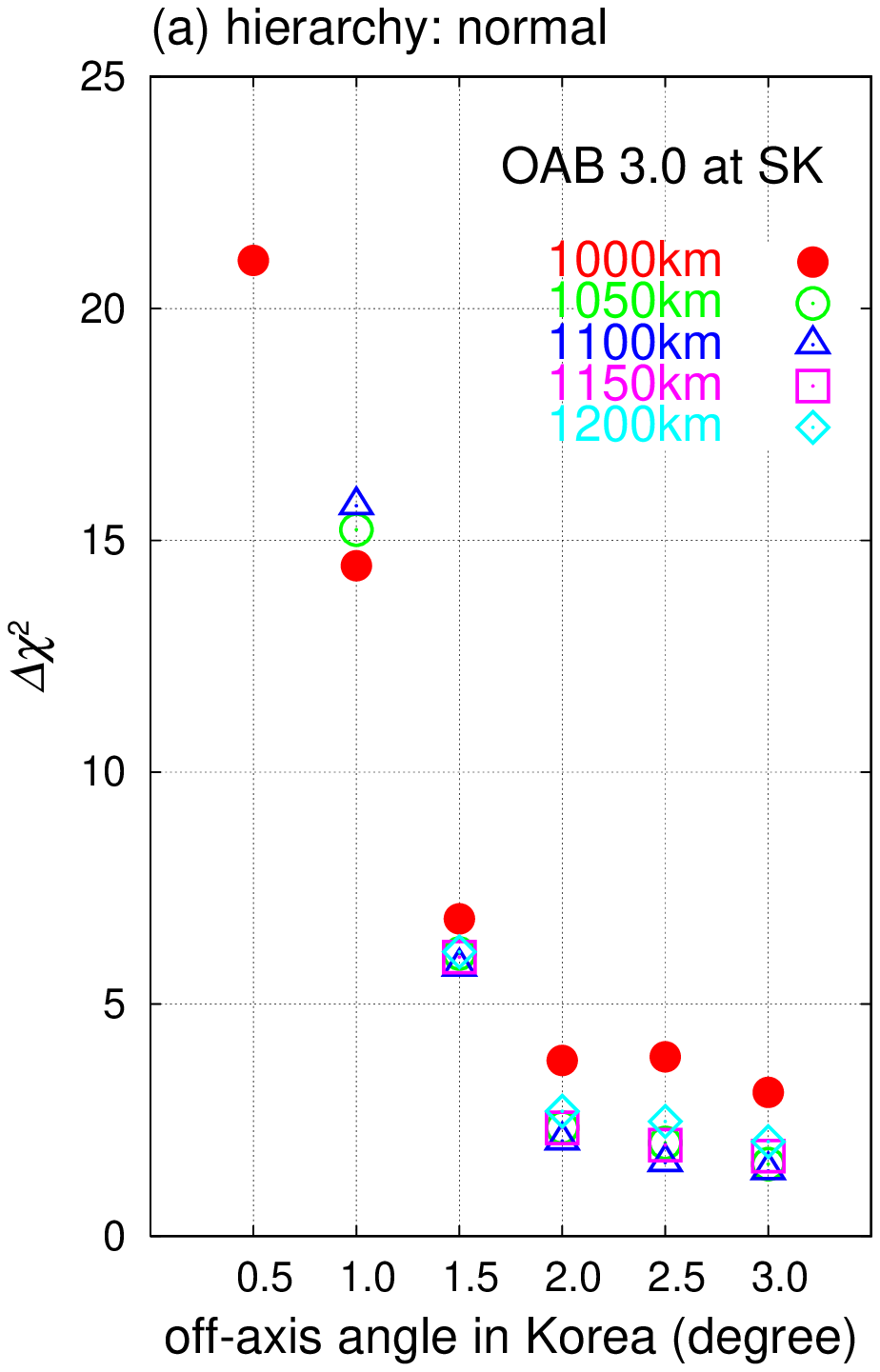}~~~~
\includegraphics[angle = 0 ,width=7.5cm]{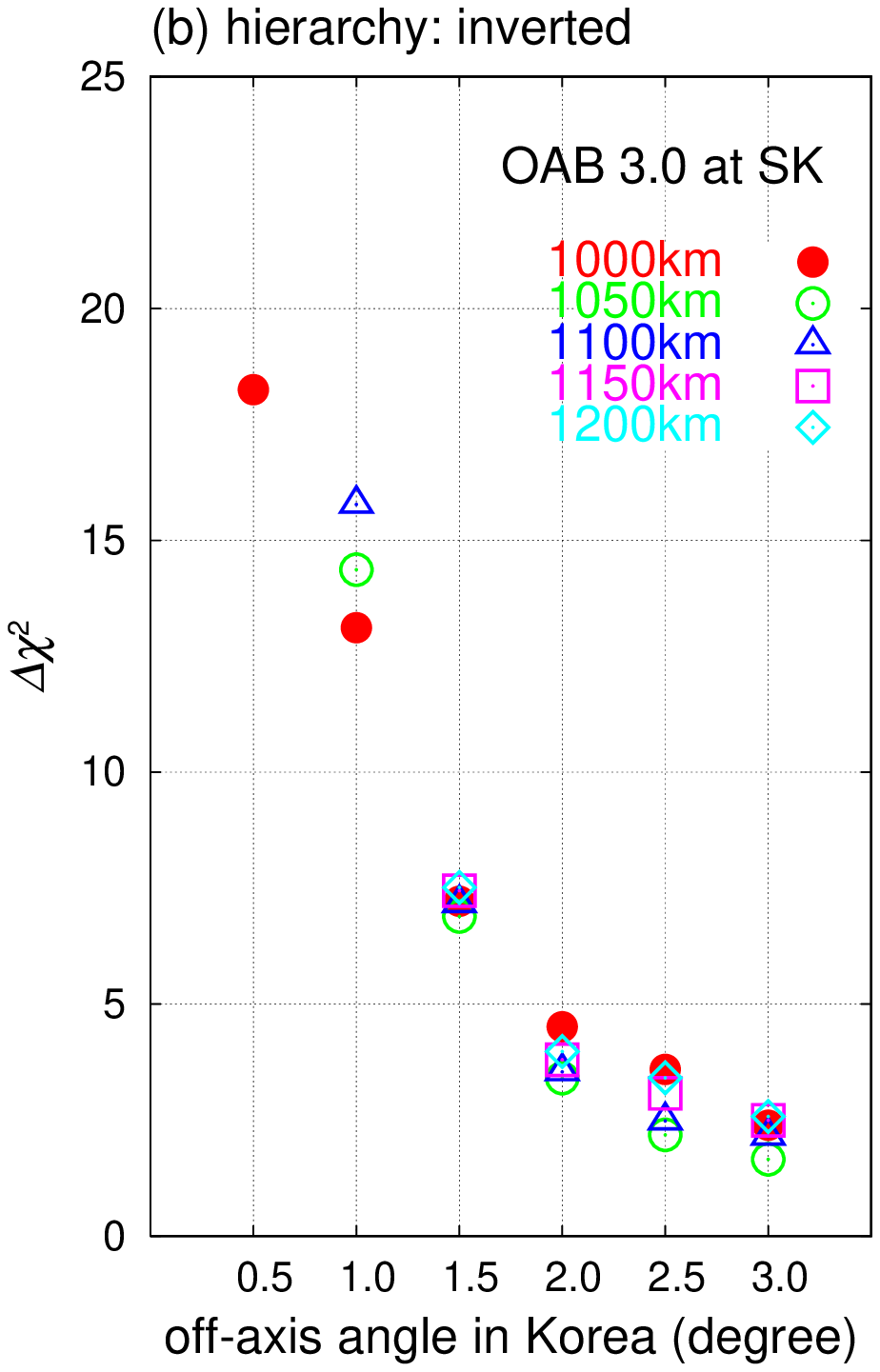}
\end{center}
\caption{
The minimum $\Delta \chi^2$ to exclude the wrong neutrino mass hierarchy
as functions of the off-axis angle and
the length of the far detector in Korea,
when the $3.0^{\circ}$ OAB reaches SK.
The left figure (a) for the normal hierarchy,
and the right figure (b) is for the inverted hierarchy. 
The input parameters are the $|m^2_3 - m^2_1| = 2.35 \times 10^{-3} {\rm eV}^2$,
$m^2_2 - m^2_1 = 8.3 \times 10^{-3} {\rm eV}^2$,
$\sin^2 \theta_{\rm \footnotesize ATM} = 0.5$,
$\sin^2 2\theta_{\rm \footnotesize SOL} = 0.85$,
$\sin^2 2\theta_{\rm \footnotesize RCT} = 0.10$ and $ \dmns = 0^{\circ}$.
$\Delta \chi^2$ is calculated for a far detector for with 100 kt
fiducial volume and for the neutrino flux with $5 \times 10^{21}$ POT
at J-PARC.
}
\label{fig:place}
\end{figure}

Since the matter profile along the Tokai-to-Korea
baseline depends on the baseline length
as shown in \figref{fig:t2korea},
we re-examine the
location dependence of the far detector capability
to determine the neutrino mass hierarchy pattern.
We show the result of our numerical calculation in \figref{fig:place},
which shows the 
minimum $\Delta \chi^2$ to reject the wrong mass hierarchy
for various combination of the 
off-axis angle and the baseline length
of Tokai-to-Korea baseline,
when the $3.0^{\circ}$ OAB reaches SK.
The left figure (a) is for the normal hierarchy
and the right figure (b) is for the inverted hierarchy. 
The input parameters are $|m^2_3 - m^2_1| = 2.35 \times 10^{-3} {\rm eV}^2$,
$m^2_2 - m^2_1 = 7.5 \times 10^{-3} {\rm eV}^2$,
$\sin^2 \theta_{\rm \footnotesize ATM} = 0.5$,
$\sin^2 2\theta_{\rm \footnotesize SOL} = 0.85$,
$\sin^2 2\theta_{\rm \footnotesize RCT} = 0.10$ and $ \dmns= 0^{\circ}$,
and $\Delta \chi^2$ is calculated for a far detector of 100 kt fiducial volume
and $5 \times 10^{21}$ POT at J-PARC.
It should be noted here that the $0.5^{\circ}$ OAB reaches Korean peninsula
only for $ L \simlt 1030$ km; see
the contour plot Fig. 1 of Ref.~\cite{t2kr-l}.
We therefore have only the $L = 1000$ km point for $0.5^{\circ}$,
three points for up to $L = 1100$ km for $1.0^{\circ}$,
all five points for higher off-axis angles in \figref{fig:place}.
\par
It is clear from the figure
that the combination of OAB $0.5^{\circ}$
at $L = 1000$ km is
most powerful to determine the mass hierarchy pattern,
for both the normal and inverted hierarchy cases.
This is because the $0.5^{\circ}$ OAB has the strongest flux 
around the first oscillation maximum \cite{t2kk-full}.
We also find that the $1.0^{\circ}$ OAB
has 
significant sensitivity to the mass hierarchy
especially at longer baseline lengths.
The value of the $\Delta \chi^2$ for $1.0^{\circ}$ OAB at $L = 1100$ km
is 15.7 (15.8)
for the normal (inverted) hierarchy, respectively,
as compared to 21.0 (18.2) for $0.5^{\circ}$ OAB at $L = 1000$ km.
On the other hand, the sensitivity of T2KK experiment on the mass hierarchy
decreases significantly for the off-axis angle greater than about
$1.5^{\circ}$.
\begin{figure}[t]
\begin{center}
\includegraphics[angle = 0 ,width=7.5cm]{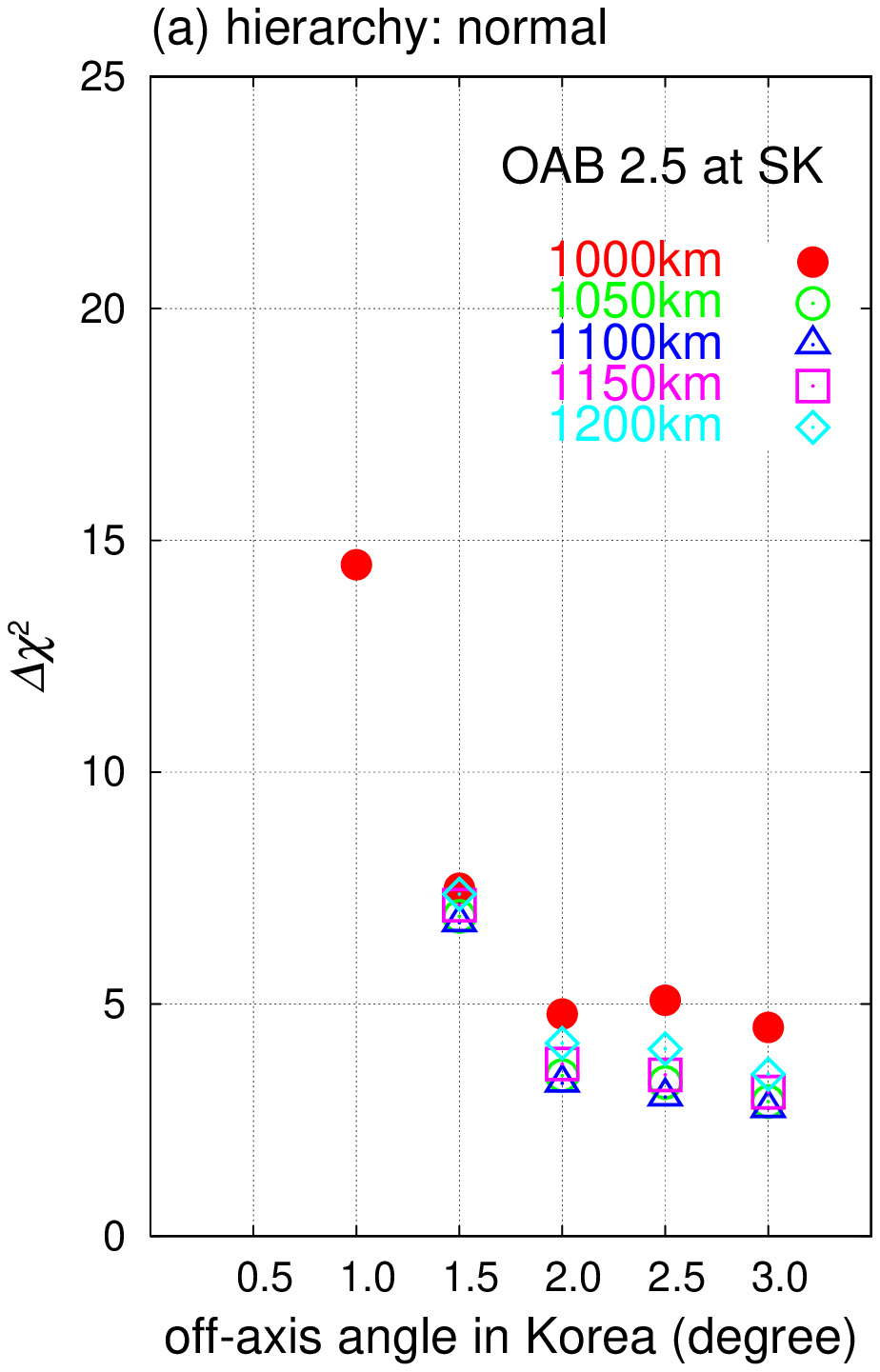}~~~~
\includegraphics[angle = 0 ,width=7.5cm]{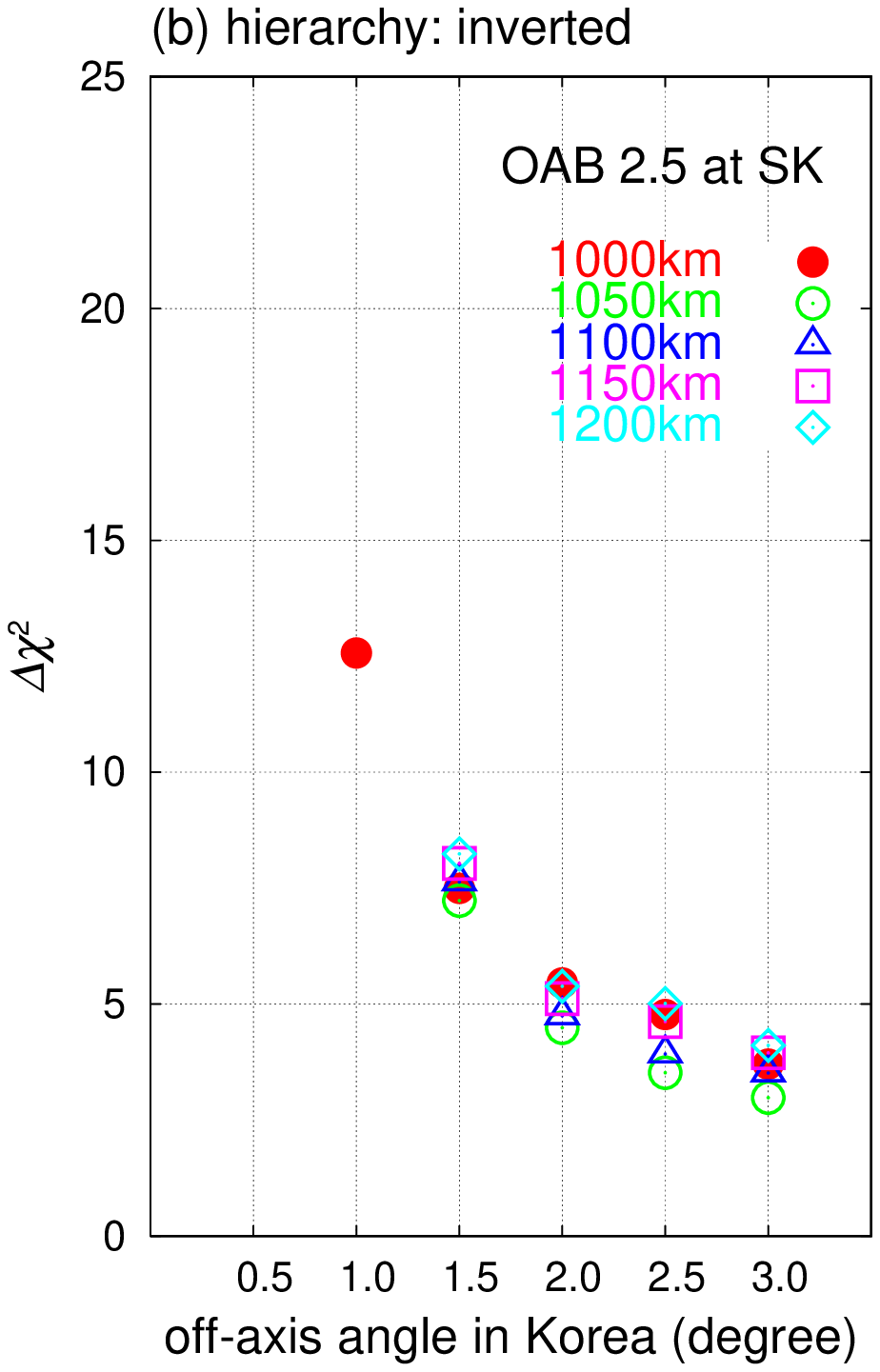}
\end{center}
\caption{The same as Fig.9 but for the case
when $2.5^{\circ}$ OAB reaches SK. 
}
\label{fig:place-i}
\end{figure}
\par
We also examin the case when the $2.5^{\circ}$ OAB reaches SK,
whose results are shown in \figref{fig:place-i}.
It is worth noting here that
no location inside Korean peninsula can
observe the neutrino beam 
at an off-axis angle $\sim 0.5^{\circ}$ :
See the off-axis angle contour plot of Fig~1
in Ref.~\cite{t2kr-l}.
The physical location of a far detector that
detects $1.0^{\circ}$ OAB in \figref{fig:place-i}
is the same as that of observing $0.5^{\circ}$
OAB in \figref{fig:place}.
In general, when we decreases 
the off-axis angle at SK by $0.5^{\circ}$,
that of a far detector in Korea increases by $\sim0 .5^{\circ}$
as can be seen from the schematic picture of \figref{fig:new-cross}.
Therefore, by comparing
the minimum $\Delta \chi^2$ values
of the corresponding points
in Fig. \ref{fig:place} and \ref{fig:place-i},
we can estimate the off-axis angle dependence
of the T2KK experiment
for each location of the far detector in Korea.
We generally find that the minimum $\Delta \chi^2$
value decreases by about $40\%$
when the off-axis angle at SK
is decreased from $3.0^{\circ}$ to $2.5^{\circ}$.
It is therefore important that
the off-axis angle at SK should be increased,
{\it i.e.} the beam
center should be oriented deeper into underground at J-PARC.
\section{Earth matter effect for the anti-neutrino beam}
In this section, we 
study the impacts of using the anti-neutrino beam
in addition to the neutrino beam in the
T2KK neutrino oscillation experiment.
\par
Potential usefulness of having both neutrino and anti-neutrino
beam run in the two detector experiments like T2KK can be
understood as follows.
The leading contributions to the $\nu_{\mu} \to \nu_e$ 
and $\bar{\nu}_{\mu} \to \bar{\nu}_e$ oscillation
shifts in the amplitude and the phase are expressed,
respectively, as
\begin{subequations}
\label{eqs:anti-difference}
\begin{eqnarray}
A^e &\sim& \pm \left( 0.11 \frac{\pi}{\Delta_{13}}
\frac{L}{295 {\rm km}} \frac{\bar{\rho}_0}{3 {\rm g/cm}^2}
- 0.29\frac{\sin \delta}{\sqrt{y}} \right)\,,
\label{eq:Ae-anti-neutrino}
\\
B^e &\sim& \mp0.08 \frac{L}{295 {\rm km}} \frac{\bar{\rho}_0}{3 {\rm g/cm}^2}
+ 0.14 \left(
\frac{\cos \delta}{\sqrt{y}} -0.11 
\right) \,,
\label{eq:Be-anti-neutrino}
\end{eqnarray}
\end{subequations}
where the upper signs are for $\nu_{\mu} \to \nu_e$ and the lower signs are
for $\bar{\nu}_{\mu} \to \bar{\nu}_e$
oscillations, and only the leading terms
in \eqref{eq:abc-e+} are kept.
Because the matter effect terms proportional to $L \bar{\rho}_0$
contribute with opposite sign for $\nu_{\mu}$ and $\bar{\nu}_{\mu}$
oscillations, we can generally expect higher sensitivity
to the mass hierarchy by comparing the oscillation probabilities
at large $L$.
In addition, since only the matter effect term
changes the sign in the phase-shift term $B^e$ of \eqref{eq:Be-anti-neutrino},
possible correlation between $\cos \delta$
and the mass hierarchy pattern can be resolved,
and we can expect improvements in the hierarchy resolving power.
On the other hand, since the $\bar{\nu}_{\mu}$ flux
is generally lower than the $\nu_{\mu}$ flux for proton synchrotron based
super beams,
and also because the background level from the secondary $\nu_e$ flux
is higher than the corresponding $\bar{\nu}_e$ background for the $\nu_{\mu} \to \nu_e$
signal, we should evaluate pros and cons quantitatively.
\par
In order to estimate the merit of using both $\nu_{\mu}$ and $\bar{\nu}_{\mu}$
beams at T2KK,
we repeat the $\Delta \chi^2$ analysis by assuming $2.5 \times 10^{21}$ POT
(2.5 years for the nominal T2K beam intensity \cite{T2K}) each for $\nu_{\mu}$ and $\bar{\nu}_{\mu}$
beams, so that the total number of
POT remains the same.
We allow independent flux normalization
errors of $3\%$ for the primary $\bar{\nu}_{\mu}$ beam,
and the secondary $\bar \nu_{e},\nu_{\mu}$ and $\nu_e$ beams.
The systematic error term of $\chi^2_{\rm sys}$ in \eqref{chi^2 define}
now depends on 14 normalization factors;
\begin{equation}
\chi^{2~{\rm (new)}}_{\rm sys} = \chi^{2~(47)}_{\rm sys}  +
 \sum_{\nu_{\alpha} = \bar{\nu}_{\mu},~\nu_{\mu},~\bar{\nu}_{e},~\nu_e}
 \left( \displaystyle\frac{\bar{f}_{\alpha} - 1}{0.03} \right)^2 \,,
 \label{eq:chi-sys-anti} 
\end{equation}
where 
$ \chi^{2~(47)}_{\rm sys}$ is defined in \eqref{chisq-sys},
and the four $\bar{f}_{\alpha}$ terms measure the flux normalization
uncertainty of the primary 
($\bar{\nu}_{\mu}$) and the secondary
($\nu_{\mu},~\bar{\nu}_{e},~\nu_e$)
fluxes in the anti-neutrino beam from J-PARC \cite{t2kk-full}.
\begin{figure}[t]
\begin{center}
\includegraphics[angle = 0 ,width=7.5cm]{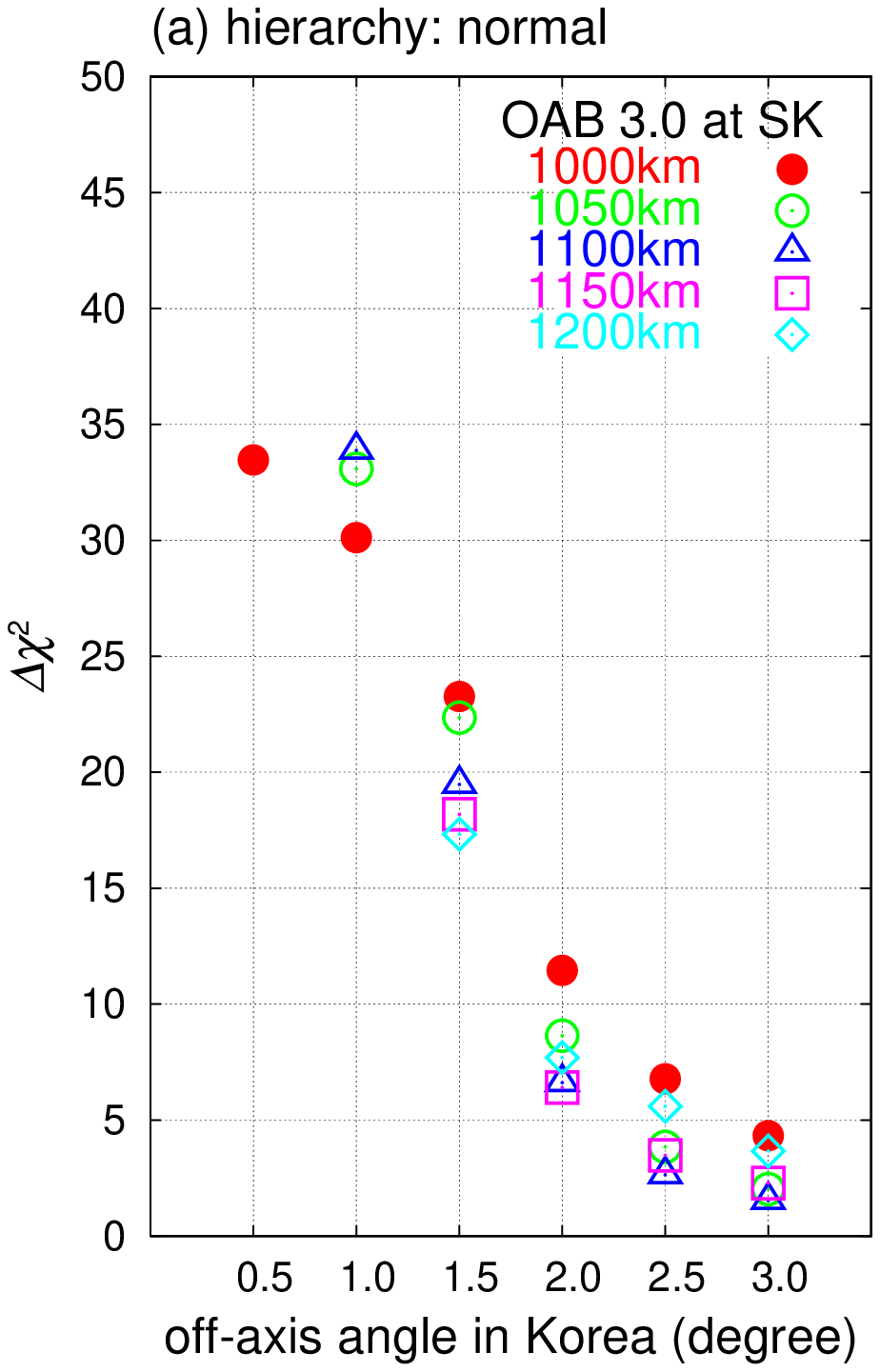}~~~~
\includegraphics[angle = 0 ,width=7.5cm]{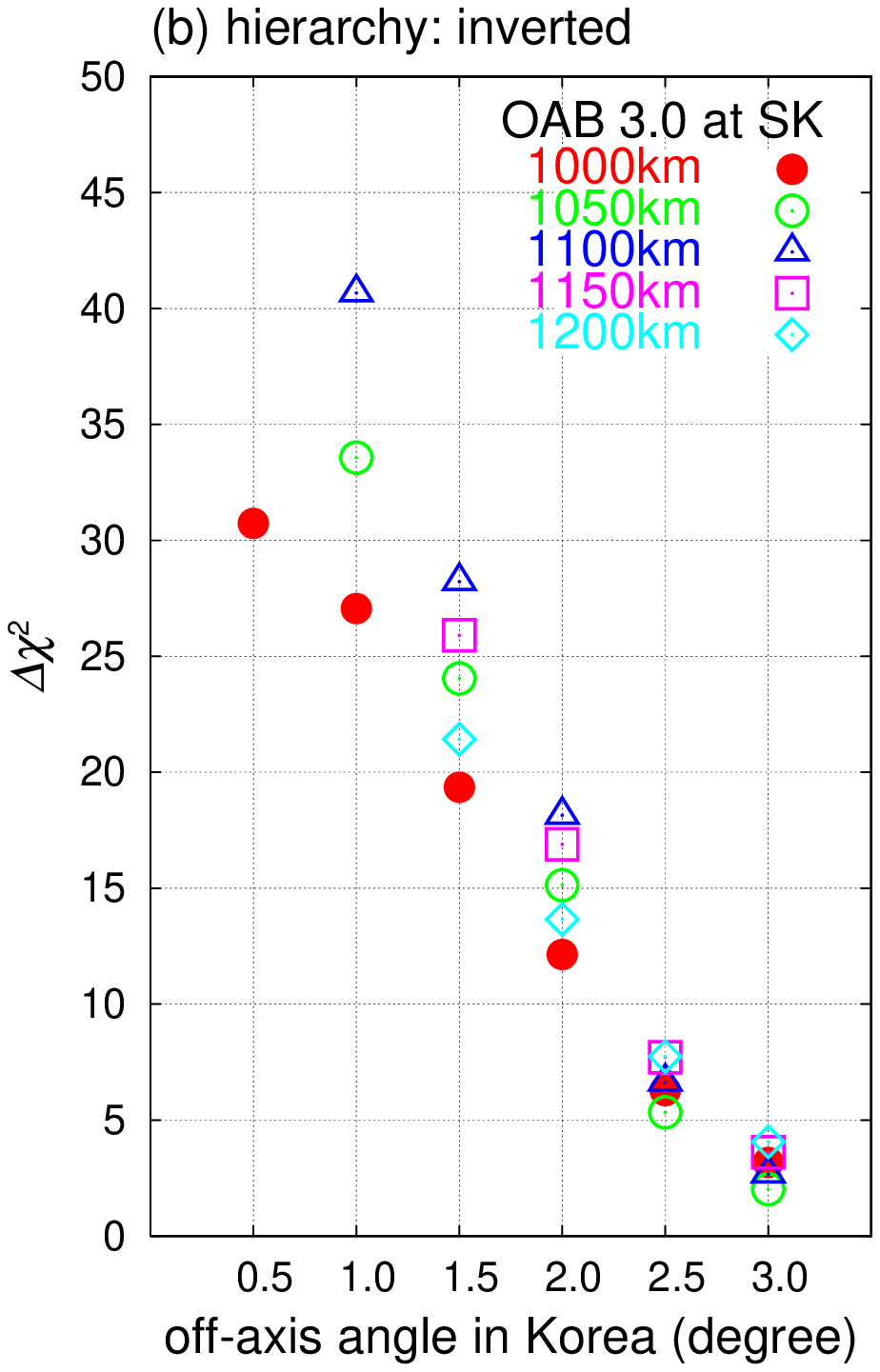}
\end{center}
\caption{
The same as \figref{fig:place} but when both neutrino and anti-neutrino beams are
used. $\Delta \chi^2$ is calculated with $2.5 \times 10^{21}$ POT
for each beam so that their sum agrees with the neutrino
flux assumed in Figs. \ref{fig:place} and \ref{fig:place-i}. 
}
\label{fig:fourie-chi-anti-normal}
\end{figure}
\begin{figure}[t]
\begin{center}
\includegraphics[angle = 0 ,width=7.5cm]{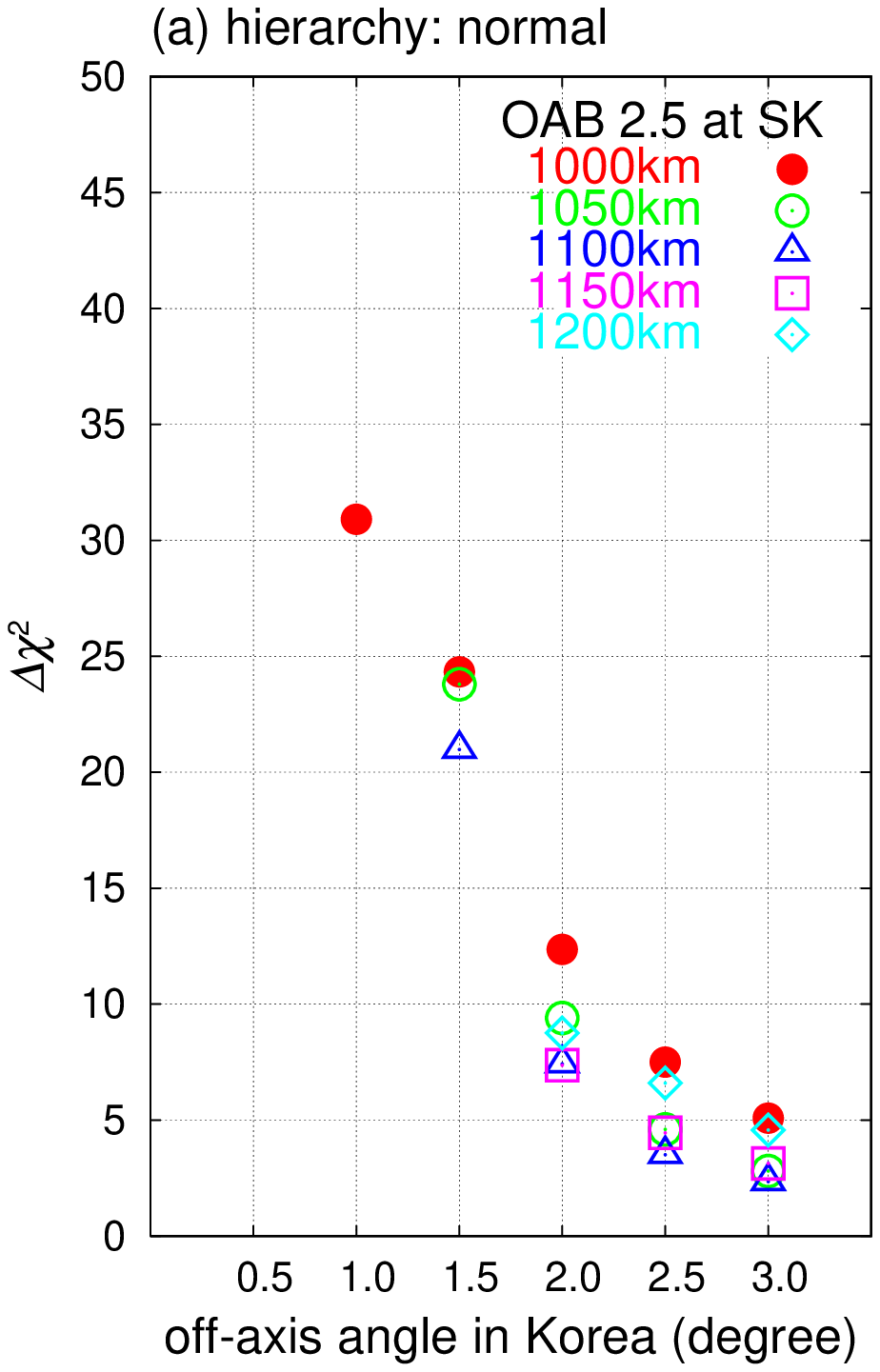}~~~~
\includegraphics[angle = 0 ,width=7.5cm]{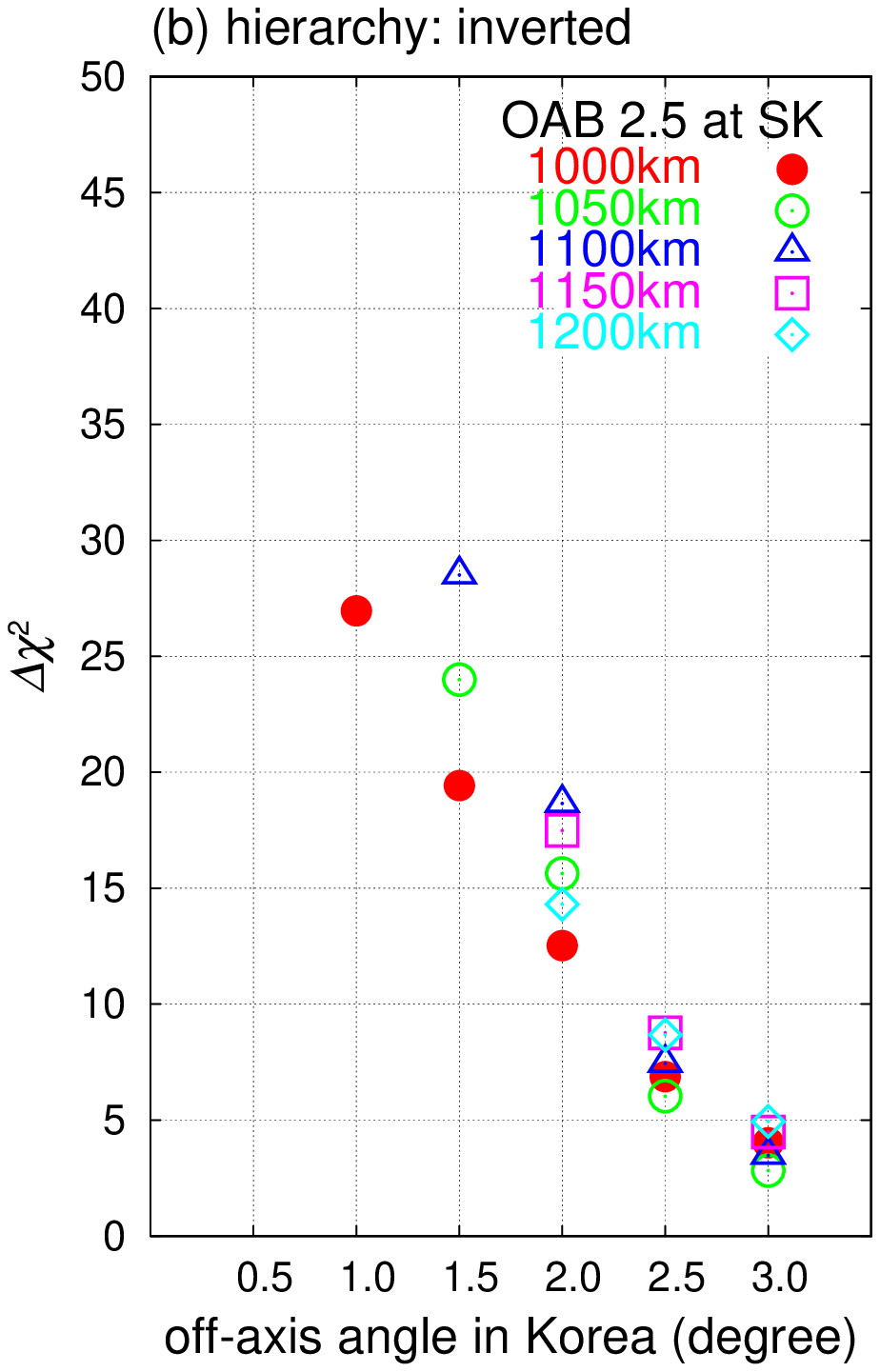}
\end{center}
\caption{The same as \figref{fig:fourie-chi-anti-normal},
but the $2.5^{\circ}$ at SK.
}
\label{fig:fourie-chi-anti-inverted}
\end{figure}
\par The results are shown in \figref{fig:fourie-chi-anti-normal}
for $3.0^{\circ}$ OAB at SK, and \figref{fig:fourie-chi-anti-inverted}
for the $2.5^{\circ}$ OAB at SK. It is remarkable that the splitting
of the total neutrino flux into half $\nu_{\mu}$ and half $\bar{\nu}_{\mu}$
results in significant improvements in the mass hierarchy resolving power
of the T2KK experiment.
For the combination of $3.0^{\circ}$ OAB at SK and
$0.5^{\circ}$ OAB at $L = 1000$ km,
the increase in $\Delta \chi^2_{\rm min}$ reads 21.0
to 33.4 for the normal hierarchy while 18.2
to 30.7 for the inverted hierarchy.
\par
It is even more striking that the
hierarchy resolving power at the 3 locations
of the far detector at $1.0^{\circ}$ OAB is as high
as that of $L = 1000$ km at 0.5 OAB in \figref{fig:fourie-chi-anti-normal}.
At $L = 1100$ km (blue open triangles),
$\Delta \chi^2_{\rm min}$  increases from 15.7 and 15.8
in \figref{fig:place}, respectively,
to 33.9 and 40.7 in \figref{fig:fourie-chi-anti-normal}
for the normal (a) and inverted (b) hierarchy case.
We find that this is partly because of our choice of the
CP phase, $\delta = 0^{\circ}$.
For $\cos \delta = 1$ and
$\sin ^2 2\theta_{\rm RCT} = 0.1$,
the phase-shift term $B^e$ in \eqref{eq:B-e+}
becomes as large as 0.4 for the $\bar{\nu}_{\mu} \to \bar{\nu}_e$ oscillation,
significantly shifting the peak location of eq.~(34)
to larger $|\Delta_{13}|$, or smaller $E_{\nu}$,
for the inverted hierarchy ($\Delta_{13} < 0$).
This shift of the oscillation maximum to lower $E_{\nu}$
compensates for the slightly softer spectrum of the $1.0^{\circ}$
flux as compared to the $0.5^{\circ}$ OAB flux.
We confirm the expected $\delta$ dependence of this
observation by repeating the analysis for $\delta = 180^{\circ}$,
where the increase in $\Delta \chi^2_{\rm min}$ is not as
dramatic as in the $\delta = 0^{\circ}$ case.
General survey of the physics capability of the T2KK experiment
over the whole parameter space of the three neutrino model
is beyond the scope of this report,
and will be reported elsewhere.
\par
We show in \figref{fig:fourie-chi-anti-inverted}
the corresponding $\Delta \chi^2_{\rm min}$ values when the
beam center is oriented upward by $0.5^{\circ}$ so that the $2.5^{\circ}$
OAB reaches SK, and hence the far detector in Korea observes
the beam at $0.5^{\circ}$ larger off-axis angles.
When we compare the $\Delta \chi^2_{\rm min}$ values in Figs.
\ref{fig:fourie-chi-anti-normal} and \ref{fig:fourie-chi-anti-inverted}
at the same far detector locations, we generally find smaller values
in \figref{fig:fourie-chi-anti-inverted}, which are consequences of the
softer neutrino beam spectra at larger off-axis angles.
As in the neutrino beam only case, it is desirable to make the off-axis
angle as large as possible at SK,
so that the far detector in Korea can observe the same beam at a smaller off-axis angle.
\section{Summary and discussion}
In this paper, we study the earth matter effects in the T2KK experiment
by using recent geophysical measurements \cite{yamato, oki, tsushima, japan-profile, korea-profile}.
\par
The mean matter density $\bar{\rho}_0$ along the Tokai-to-Kamioka baseline is found to be $2.6 \density$.
The presence of Fossa Magna along the T2K baseline makes the matter density
distribution concave, or Re($\bar{\rho}_1) > 0$ for the
first coefficient of the Fourier expansion of the density distribution
along the baseline, which is opposite from what is naively expected from
the spherically symmetric model of the earth matter distribution
such as PREM \cite{prem}.
We find that both the reduction of the average matter density
$\bar{\rho}_0$ and the positive Re$(\bar{\rho}_{1})$ contribute
positively to the mass hierarchy resolving power 
of the T2KK experiment, because the sensitivity grows as the
difference in the magnitude of the earth matter effect between
the oscillation probabilities observed at the near (SK) and a far (Korea) detectors.
\par
As for the Tokai-to-Korea baselines,
we find that the average matter density $\bar{\rho}_0$
grows from 2.85 g/cm$^3$ at $L = 1000$ km to 2.98 g/cm$^3$ at $L = 1100$ km,
which are significantly higher than the PREM value
of 2.70 g/cm$^3$ ($L$ = 1000 km) and 2.78 g/cm$^3$ ($L$ = 1100 km);
see \tableref{tab:compare-prem}.
This is essentially because both the Conrad discontinuity
between the upper and the lower crust, and the Moho-discontinuity
between the crust and the mantle are significantly higher than PREM
in the main region of the baseline below the Japan/East sea.
This also contributes positively to the mass
hierarchy measurement. The distribution is convex as in PREM
but Re$\bar{\rho}_1/\bar{\rho}_0$ is smaller in magnitude
mainly becomes of the thin crust under the sea. 
One subtle point which we notice is that the $ L = 1000$ km
baseline almost touches the Moho-discontinuity 
at its bottom, see \figref{fig:t2korea},
and hence small error in the depth of the discontinuity can change
the average density significantly.
If the Moho-discontinuity at about 20 km below the sea level is shallower by 0.7 km,
which is the present error of the seismological measurement,
$\bar{\rho}_0$ grows from 2.85 g/cm$^3$ to 2.90 g/cm$^3$,
by about 2$\%$. On the other hand,
this uncertainty does not affect the T2KK physics capability
significantly because
the major error in the earth matter density enters
when the sound velocity data is converted to the matter density,
for which we assign $6\%$ error in our analysis; see \figref{fig:v-rho-trans}.
\par
With the updated matter density distributions,
we repeat the $\Delta \chi^2$ analysis of Refs. \cite{t2kr-l, t2kk-full},
which estimates the mass hierarchy
resolving power of the T2KK experiment, by performing
a thought experiment with a far detector
of 100 kt fiducial volume in various location of Korea along the T2K beam line.
In order to perform the full parameter scan effectively, we introduce a new
algorithm to compute the oscillation probabilities exactly
for step-function-like matter density profile,
where the earth matter effect is computed by diagonalizing real
symmetric $3\times3$ matrix by decoupling it from the evolution
due to the $\theta_{23}$ mixing and the CP phase \cite{hagiwara-okamura}.
Analytic and semi-analytic approximations for the transition and survival probabilities
are introduced to obtain physical interpretations of the numerical results.
\par
We confirm the previous findings of Refs.~\cite{t2kr-l, t2kk-full} that
the mass hierarchy resolving capability is maximized when the T2K
beam is oriented downward to $3.0^{\circ}$ OAB
at SK and the far detector is placed in the south-east coast
of Korean peninsula
at $L \sim 1000$ km where $\sim 0.5^{\circ}$ OAB can be observed.
In this report, we repeat
the analysis by splitting the total beam time
into half neutrino ($\nu_{\mu}$) and half anti-neutrino ($\bar{\nu}_{\mu}$) beams,
and find significant improvements in the mass hierarchy resolving power.
Most remarkably, we find that the highest level of sensitivity can now
be achieved for a far detector in locations
where the T2K neutrino beam can be detected
at an off-axis angle up to  $\sim 1^{\circ}$,
because of the significant difference in the $E_{\nu}$-dependence of the oscillation
probability between $\nu_{\mu} \to \nu_e$ and $\bar{\nu}_{\mu} \to \bar{\nu}_e$.
This observation expands significantly the area in which a far detector
can be most effective in determining the neutrino mass hierarchy pattern.
\vspace{0.1cm}
\\
{\it Acknowledgments}
\vspace{0.1cm}
\\
We thank our experimentalist colleagues
Y.~Hayato,
A.K.~Ichikawa,
T.~Kobayashi
and
T.~Nakaya,
from whom we learn about the K2K and T2K experiments.
We thank N.~Isezaki and M.~Komazawa for teaching us about the geophysics
measurements in Japan/East sea.
We are also grateful to 
T.~Kiwanami and
M.~Koike
for useful discussions and comments.
We thank the Aspen Center for Physics
and the Phenomenology Institute at the University of Wisconsin
for hospitality.
The work is supported in part
by the Grant in Aid for scientific research ($\#$20039014)
from MEXT, and in part by
the Core University Program of JSPS.
The numerical calculations were carried out on
KEKCC at KEK and
Altix3700 BX2 at YITP in Kyoto University.
\par

\setcounter{section}{0} 
\renewcommand{\thesection}{\Alph{section}} 
\setcounter{equation}{0} 
\renewcommand{\theequation}{A.\arabic{equation}} 

\section*{Appendix}
\hspace{11pt}
In the appendix, we present the second order perturbation formula
of the time-evolution operator
for the location $(x)$ dependent Hamiltonian of \eqref{eq:hamiltonian-all}:
\begin{eqnarray}
H(x) &=& H_0 + \overline{H}_1 + \delta H_1(x) \,,
\label{apeq:hamiltonian}
\end{eqnarray}
where each term has the matrix element
\begin{subequations}
\label{eqs:matrix-element}
\begin{eqnarray}
\left \langle \nu_{\beta} \right |
H_0
\left| \nu_{\alpha} \right \rangle
&=&
\frac{m^2_3 - m^2_1}{2E}
U_{\beta 3}U^*_{\alpha 3}
\,,
\label{eq:matrix-element-h0}
\\
\left \langle \nu_{\beta} \right |
\overline{H}_1
\left| \nu_{\alpha} \right \rangle
&=&
\frac{m^2_2 - m^2_1}{2E}
U_{\beta 2}U^*_{\alpha 2}
+
\frac{\bar{a}_0}{2E}
\delta_{\beta e}
\delta_{\alpha e}
\,,
\label{eq:matrix-element-h1bar}
\\
\left \langle \nu_{\beta} \right |
\delta H_1(x)
\left| \nu_{\alpha} \right \rangle
&=&
\frac{1}{E}
\sum^{\infty}_{k=1}
\left[
{\rm Re}(\bar{a}_k) \cos \left(\frac{2 \pi k x}{L}\right)
+{\rm Im}(\bar{a}_k) \sin \left(\frac{2 \pi k x}{L}\right)
\right]
\delta_{\beta e} \delta_{\alpha e}\,.
\label{eq:matrix-element-deltah1}
\end{eqnarray}
\end{subequations}
The first two terms, $H_0$ and $\overline{H}_1$,
do not depend on the location $x$, whereas in
the last term $\delta H_1(x)$
the $x$-dependent part of the matter density distribution along the baseline,
$\rho(x) - \bar{\rho}_0$ is expressed as a Fourier expansion.
The second term in \eqref{eq:matrix-element-h1bar} gives
the matter effect due to the average matter density $\bar{\rho}_0$.
\par
In the leading order, we consider only $H_0$
of eq.~(\ref{eq:matrix-element-h0}).
The time evolution via $H_0$ can readily be solved 
and the $\left | \nu_{\alpha} \right \rangle \to \left | \nu_{\beta} \right \rangle$
transition matrix element is
\begin{eqnarray}
S_0(L)_{\beta \alpha} &=&
\left \langle \nu_{\beta} \right |
e^{-iH_0 L}
\left| \nu_{\alpha} \right \rangle
=
U_{\beta1}U_{\alpha 1}^* + U_{\beta2}U_{\alpha 2}^* + U_{\beta3}U_{\alpha 3}^* e^{-i\Delta_{13}}
\nonumber \\
&=& \delta_{\beta \alpha} + U_{\beta 3}U_{\alpha 3}^*
\left(
e^{-i \Delta_{13}} -1
\right) \,,
\label{eq-s0-ab}
\end{eqnarray}
with $\Delta_{13}$ as in \eqref{eq:Deltaij}.
In the next order, we divide the corrections
into two parts,
\begin{eqnarray}
S_1(L)_{\beta \alpha} &=& S_1^{(1)}(L)_{\beta \alpha} + S_1^{(2)}(L)_{\beta \alpha} \,,
\label{eq:div-s1}
\end{eqnarray}
where the first term is linear in $\bar{H}_1$;
\begin{eqnarray}
S_1^{(1)}(L)_{\beta \alpha}
&=& -i \left \langle \nu_{\beta} \right |
\left(\int^{L}_{0} dx e^{iH_0 (x-L)} \bar{H}_1
e^{-iH_0 x}\right)
\left| \nu_{\alpha} \right \rangle
\nonumber \\
&=&
-i
\Delta_{12} U_{\beta 2} U_{\alpha 2}^*
-i \frac{\bar{a}_0 L}{2E}
\biggl[
 \delta_{\beta e} \delta_{\alpha e}
- A_{\beta \alpha}
+ 2 B_{\beta \alpha}
\biggr.
\nonumber \\
&&
+
\left.
\left( e^{-i \Delta_{13} } -1 \right)
\left\{ B_{\beta \alpha}
+\frac{i}{\Delta_{13}}
\left(A_{\beta \alpha}
-2 B_{\beta \alpha}
\right)
\right\}
\right] \,,
\label{eq:calc-s1-1}
\end{eqnarray}
and the second term is linear in $\delta H_1$.
\begin{eqnarray}
S_1^{(2)}(L)_{\beta \alpha} &=& -i
\left \langle \nu_{\beta} \right |
\left(\int^{L}_{0} dx e^{iH_0 (x-L)} \delta H_1(x)
e^{- iH_0 x}
\right)
\left| \nu_{\alpha} \right \rangle
\nonumber \\
&=&
\frac{L}{E}
\frac{ e^{-i \Delta_{13} } -1}{\Delta_{13}}
\times
\nonumber \\
&&
\sum^{\infty}_{k = 1}
\left[
\frac{{\rm Re} (\bar{a}_k) \Delta_{13}^2 }{\Delta_{13}^2 - \left( 2 \pi k \right)^2} 
\left( A_{\beta \alpha} - 2 B_{\beta \alpha} \right)
- i
\frac{{\rm Im} (\bar{a}_k)  2 \pi k \Delta_{13} }{\Delta_{13}^2 - \left( 2 \pi k \right)^2} 
C_{\beta \alpha}
\right] 
\,.
\label{eq:calc-s1-2}
\end{eqnarray}
Here the coefficients $A_{\beta \alpha}$, $B_{\beta \alpha}$, and $C_{\beta \alpha}$
are quadratic terms of the MNS matrix elements,
\begin{subequations}
\begin{eqnarray}
A_{\beta\alpha} &\equiv& 
\delta_{\beta e}U_{e3} U_{\alpha3}^*
+ U_{\beta 3} U_{e3}^* \delta_{\alpha e}
\,, \label{eq:def-A1} \\
B_{\beta\alpha} &\equiv& 
U_{\beta 3}U_{\alpha 3}^{*} |U_{e3}|^2
\,, \label{eq:def-B1} \\
C_{\beta\alpha} &\equiv& 
  \delta_{\beta e}U_{e3} U_{\alpha3}^*
- U_{\beta 3} U_{e3}^* \delta_{\alpha e}
\,. \label{eq:def-C1}
\end{eqnarray}
\end{subequations}
Generally, the $B_{\beta \alpha}$ terms are much smaller than
the other terms because of the constraint $|U_{e3}|^2 < 0.044$.
Since $A_{\mu \mu} = 0$ and $|B_{\mu \mu}| \sim |U_{\mu 3} U_{e3}|^2 \ll 1$,
we can approximate $ S^{(1)}_1 (L)_{\mu \mu}$ rather accurately  
by $-i \Delta_{12} |U_{\mu 2}|^2$, which does not depend on
the matter effect term $\bar{a}_0$.
On the other hand,
$|A_{e \mu}|$ can be as large as
$|U_{\mu 2}U_{e2}|$ in $S_1^{(1)}(L)_{e\mu}$, and
the matter effect can be significant.
\par
As for the $x$-dependent perturbation $S^{(2)}_{\beta \alpha}$,
we note that the terms proportional to Re$(\bar{a}_k)$ are in 
phase with the leading term of \eqref{eq-s0-ab}
for $\beta \alpha = e \mu$, whereas those with Im$(\bar{a}_k)$
are orthogonal. Therefore the Im$(\bar{a}_k)$ terms contribute to
the $\nu_{\mu} \to \nu_{e}$ oscillation probability only as 
Im$(\bar{a}_k)^2 |C_{e \mu}|^2 \sim (\bar{a}_k)^2 |U_{e3}|^2$,
which is strongly suppressed. 
Contributions to the $\nu_{\mu} \to \nu_{\mu}$ survival probability are
strongly suppressed again in this order due to the smallness
of $|A_{\mu\mu} - 2B_{\mu\mu}|$ and $C_{\mu \mu} = 0$.
\par
Finally, the second order correction, $S_2$,
is evaluated only for $x$-independent perturbation, $\overline{H}_1$,
because the magnitude of all the non-zero Fourier coefficients are
significantly smaller than the average, $\bar{\rho}_0$;
see Figs.~\ref{fig:fourie-295}, \ref{fig:f-real} and \tableref{tab:compare-prem}.
We find
\begin{eqnarray}
S_2(L)_{\beta \alpha} &=& -
\left \langle \nu_{\beta} \right |
\left( \int^{L}_{0} dx \int^{x}_{0} dy 
e^{iH_0 (x-L)} \overline{H}_1
e^{iH_0 (y-x)} \overline{H}_1 e^{-iH_0 y} \right)
\left| \nu_{\alpha} \right \rangle
\\
&=&-i \sum_{\gamma}
\int^{L}_{0}dx
\left[\frac{\Delta_{12}}{L}U_{\beta 2}U_{\gamma 2}^*
+ \frac{\bar{a}_0}{2E}\left\{
\delta_{\beta e}\delta_{e \gamma}
+ U_{\beta 3}U_{e 3}^* \delta_{e \gamma}
\left(
e^{i\frac{\Delta_{13}}{L}(x-L)} -1
\right)
\right\}
\right]
S_1^{(1)}(x)_{\gamma \alpha} \nonumber \\
&=&
-\frac{\Delta_{12}^2}{2} U_{\beta 2} U_{\alpha 2}^*
\nonumber \\
&&- \frac{\Delta_{12}}{2} \frac{\bar{a}_0L}{2E}
\left[
D_{\beta \alpha}
  -\left\{ 1 + \frac{2i}{\Delta_{13}}
+\frac{2\left(e^{-i \Delta_{13}} -1\right)}{\Delta_{13}^2}
   \right\}E_{\beta \alpha}
\right]
\nonumber \\
&&
-\left(\frac{\bar{a}_0L}{2E}\right)^2\left[
 \frac{1}{2}
  \left\{
    \left( 1 - |U_{e3}|^2 \right)
    \left(\delta_{\beta e}\delta_{\alpha e}
      -A_{\beta \alpha}
    \right)
    + B_{\beta \alpha}
 \right\}
\right.
\nonumber \\
&&\hspace{1.1cm}
\left.
 - \frac{i}{\Delta_{13}}
 \left\{ \delta_{\beta e} \delta_{\alpha e} |U_{e3}|^2
         + \left( 1 - 3 |U_{e3}|^2\right)
        A_{\beta \alpha} - 3\left(1 - 2|U_{e3}|^2 \right) B_{\beta \alpha} 
   \right\}
\right.
\nonumber 
\\
&&\hspace{1.1cm}
\left.
 +\left( e^{-i\Delta_{13}} - 1 \right)
  \left\{ \frac{1}{2}|U_{e3}|^2 B_{\beta \alpha} + \frac{i}{\Delta_{13}}
          \left( |U_{e3}|^2 A_{\beta \alpha} + \left( 1 - 3 |U_{e3}|^2\right)B_{\beta \alpha}\right)
  \right.       
\right.
\nonumber\\
&&\hspace{1.1cm}
\left.
 \left.
 +\frac{1}{\Delta_{13}^2}
   \left( - \delta_{\beta e}\delta_{\alpha e}|U_{e3}|^2
   - \left( 1 - 3|U_{e3}|^2 \right) A_{\beta \alpha}
  +3\left(1 - 2|U_{e3}|^2 \right) B_{\beta \alpha} 
    \right)
 \right\}
\right] \,,
\end{eqnarray}
where $D_{\beta \alpha}$ and $E_{\beta \alpha}$ are the
MNS matrix element factors
\begin{subequations}
\begin{eqnarray}
D_{\beta \alpha} &=& \delta_{\beta e}U_{e2}U_{\alpha 2}^* + U_{\beta 2}U_{e2}^* \delta_{\alpha e} \,, \\
E_{\beta \alpha} &=& U_{\beta2}U_{e2}^*U_{e3}U_{\alpha3}^*+ U_{\beta3}U_{e3}^*U_{e2}U_{\alpha2}^*\,.
\end{eqnarray}
\end{subequations}
It should be noted that although
$\Delta_{12}^2$ is small, $0.01$, around the first oscillation maximum $|\Delta_{13}| \sim \pi$,
it is significant around the second oscillation maximum $|\Delta_{13}| \sim 3 \pi$
where it grows to $0.1$.
The combination, $\Delta_{12} (\bar{a}_0L/2E)$
is about 0.02 for T2K and 0.06 for $L \sim 1000$ km
around $|\Delta_{13}| \sim \pi$.
The matter effect term squared, $(\bar{a}_0L/2E)^2$,
is 0.03 at Kamioka while it grows to $\sim 0.3$ or larger for the
Tokai-to-Korea baselines.

\end{document}